\documentclass[]{aa}
\usepackage{epsf,palatino,url,graphicx,wrapfig,array,setspace,amsmath,amssymb,fancyhdr,multirow,lscape,appendix,rotating,txfonts}
\usepackage{graphics,epsfig,txfonts}
\usepackage{natbib}
\usepackage{subcaption}
\bibpunct{(}{)}{;}{a}{}{,}

\usepackage{graphicx}
\usepackage[flushleft]{threeparttable}
\usepackage{booktabs} 

\usepackage{longtable}
\usepackage{amssymb}
\usepackage{xcolor}
\usepackage{color}
\usepackage{lipsum}
\usepackage{textcomp}
\usepackage[colorlinks=true,linkcolor=blue,citecolor=blue,
bookmarks=true,bookmarksopen=true,bookmarksnumbered=true,draft=false]{hyperref}
\usepackage{enumerate}

\usepackage[normalem]{ulem}
\usepackage{soul}
\usepackage[switch]{lineno}
\usepackage{blindtext}

\usepackage{epsf}
\usepackage{amssymb}
\usepackage{epsfig}
\usepackage{pdflscape}
\usepackage{amsmath}
\usepackage{graphicx}

\newcommand{\oergs}[1]{$10^{#1}$ erg s$^{-1}$}

\newcommand{\ltsima}{$\buildrel < \over {\sim}$}
\newcommand{\lsim}{\lower.5ex\hbox{\ltsima}}
\newcommand{\gtsima}{$\buildrel > \over {\sim}$}
\newcommand{\gsim}{\lower.5ex\hbox{\gtsima}}

\newcommand{\swift}{{\it Swift}\xspace}

\newcommand{\xmm}{{\it XMM-Newton}\xspace}

\newcommand{\nus}{{\it NuSTAR}\xspace}
\newcommand{\f}{$f_{\rm col}$\xspace}

\begin{document}

\title{ULX spectra revisited: Accreting, highly magnetized neutron stars as the engines of ultraluminous X-ray sources}

\author{Filippos Koliopanos\inst{1,2}\thanks{\email{fkoliopanos@irap.omp.eu}}
          \and  Georgios Vasilopoulos\inst{3}
          \and  Olivier Godet\inst{1,2}
          \and  Matteo Bachetti\inst{4}
          \and  Natalie A. Webb\inst{1,2}
          \and  Didier Barret\inst{1,2}  }

\titlerunning{ULX}
\authorrunning{Koliopanos et al.}

\institute{CNRS, IRAP, 9 Av. colonel Roche, BP 44346, F-31028 Toulouse cedex 4, France
          \and Universit{\'e} de Toulouse; UPS-OMP; IRAP, Toulouse, France  
          \and Max-Planck-Institut für Extraterrestrische Physik, Giessenbachstraße, 85748 Garching, Germany
          \and INAF/Osservatorio Astronomico di Cagliari, via della Scienza 5, I-09047 Selargius (CA), Italy}

\date{Received ?? ??? 2016 / Accepted ?? ??? 2016}

  \abstract
   {}
   {In light of recent discoveries of pulsating ultraluminous X-ray sources (ULXs) and recently introduced theoretical schemes that propose neutron stars (NSs)
   as the central engines of ULXs, we revisit the spectra of eighteen well known ULXs, in search of indications that favour this newly emerging hypothesis.}
   {We examine the spectra from high-quality \xmm and  \nus observations. We use a combination of elementary black body and multicolour disk black body (MCD) models, to diagnose the predictions of classic and novel theoretical models of accretion onto NSs. We re-interpret the well established spectral characteristics of ULXs in terms of accretion onto lowly or highly magnetised NSs, and explore the resulting parameter space for consistency.}
   {We confirm the previously noted presence of the low-energy (${{\lesssim}}6$\,keV)
   spectral rollover and argue that it could be interpreted as due to thermal emission. 
   The spectra are well described by a double thermal model consisting of a ``hot'' (${{\gtrsim}}1\,$keV) and a ``cool'' (${{\lesssim}}0.7\,$keV) multicolour black body (MCB). Under the assumption that the ``cool'' MCD emission originates in a disk truncated at the neutron star magnetosphere, we find that all ULXs in our sample are consistent with accretion onto a highly magnetised ($B{{\gtrsim}}10^{12}$\,G) neutron star. We note a strong correlation between the strength of the magnetic field, the temperature of the ``hot'' thermal component and the total unabsorbed luminosity. Examination of the \nus data supports this interpretation and also confirms the presence of a weak, high-energy (${\gtrsim}15$\,keV) tail, most likely the result of modification of the MCB emission by inverse Compton scattering.  
   We also note that the apparent high-energy tail, may simply be the result of mismodelling of MCB emission with an atypical temperature ($T$) versus radius ($r$) gradient, using a standard MCD model with a fixed gradient of $T{{\sim}}r^{-0.75}$.
   }
   {We have offered a new and robust physical interpretation for the  dual-thermal spectra of ULXs. We find that the best-fit derived parameters of our model, are in excellent agreement with recent theoretical predictions that favour super-critically accreting NSs as the  engines of a large fraction of ULXs.  Nevertheless, the considerable degeneracy between models and the lack of unequivocal evidence cannot rule out other equally plausible interpretations. Deeper broadband observations and time-resolved spectroscopy are warranted to further explore this newly emerging framework.}
   {}

\maketitle

\section{Introduction}
\label{sec-intro}
Ultraluminous X-ray sources (ULXs) are off-nuclear, extragalactic X-ray sources with isotropic luminosities that exceed the Eddington limit for a stellar-mass black hole  (BH) ($M_{\rm BH}{{\lesssim}}20\,M_{\odot}$: see \citealt{2011NewAR..55..166F} and references therein, and also \citealt{2016AN....337..349B} and \citealt{2017ARA&A..55..303K} for up-to-date and comprehensive reviews). It was initially suggested that ULXs were rare instances of intermediate-mass BHs accreting at sub-Eddington rates \citep{1999ApJ...519...89C,2000ApJ...535..632M,2001MNRAS.321L..29K,2003ApJ...585L..37M}; essentially a scaled-up version of standard galactic BH X-ray binaries (BH-XRBs). However, it was quickly realised that a considerable fraction (if not all) of the ULX population can be powered by stellar-mass BHs accreting at super-Eddington rates \citep[e.g.][]{2003ApJ...596L.171G,2004NuPhS.132..369G,2004MNRAS.349.1193R,2007MNRAS.377.1187P,2009MNRAS.393L..41K}. Furthermore, the astounding discovery of a pulsating ULX (PULX: \citealt{2014Natur.514..202B}) demonstrated that ULXs can be powered by neutron stars (NSs). After the initial discovery by \citeauthor{2014Natur.514..202B}, two more PULXs have been detected \citep{2017Sci...355..817I,2016ApJ...831L..14F}. 

The discovery of NS-ULXs provided further support to the scenario of super-Eddington accretion onto a stellar-mass compact object as the power source of ULXs, but also posed the crucial question of the potential prevalence of NSs as the engines of ULXs.  Indeed, it has been known for some time that most of the ULXs do not transition through the phenomenological states of standard BH-XRBs, that is,~spectral transitions between hard and soft states and the appearance and quenching of relativistic jets \citep[for a review of spectral and temporal properties of NS-  and BH-XRBs see e.g.][]{2001AdSpR..28..307B,2006ARA&A..44...49R,2007A&ARv..15....1D,2010LNP...794...17G}.  While some ULXs exhibit significant luminosity and spectral variations \citep[e.g.][]{2004ApJS..154..519S,2013MNRAS.435.1758S,2016ApJ...831..117F} , they do not transition between the markedly different and characteristic ``hard'' and ``soft'' spectral states of nominal, sub-Eddington BH-XRBs \cite[e.g.][]{2008ApJ...687..471B,2010ApJ...724L.148G,2013MNRAS.435.1758S}. Interestingly two hyperluminous X-ray sources (HLXs) --  which have luminosities exceeding ${\sim}$\oergs{41} and are strong IMBH candidates -- seem to follow the transition pattern of stellar-mass BH-XRBs (ESO~243-49~HLX-1: \citealt{2009ApJ...705L.109G}; \citealt{2011ApJ...743....6S}; \citealt{2012Sci...337..554W}, or M82~X-1: \citealt{2010ApJ...712L.169F}). In this work we only consider ULXs that are limited to luminosities ${\lesssim}10^{41}$\,erg/s.

In addition to the lack of state transitions, the spectra of numerous ULXs are significantly different from the typical spectra of BH-XRBs. More specifically, the majority of ULX spectra feature a notable spectral curvature above ${{\sim}}6\,$keV \cite[e.g.][]{2006MNRAS.368..397S,2009MNRAS.397.1836G,2013MNRAS.435.1758S}. It has been proposed that perhaps the unusual spectra of ULXs correspond to a novel state of super-critical accretion, dubbed the {\it ultralumimous state} \citep{2009MNRAS.397.1836G}. In subsequent works it was posited that ULX spectra can be empirically classified into three classes based on their spectral shape: singly peaked {\it broadened disc} (BD) class and two-component hard ultraluminous (HUL) and soft ultraluminous (SUL) states \citep{2013MNRAS.435.1758S}. Recognising the physical mechanisms underlying the observed spectral characteristics of ULXs is a crucial step towards understanding accretion at super-Eddington states but also determining the nature of their accretor. 

In the recent months leading to this publication, an increasing number of compelling theoretical considerations -- that point to NSs as plausible engines behind ULX emission -- have been put forward by numerous authors  \citep[e.g.][]{2016MNRAS.458L..10K,2017MNRAS.468L..59K,2017MNRAS.tmp..143M}.
Motivated by these findings and the apparent spectral and temporal similarities between ULXs and NS-XRBs we decided to revisit the spectra of known ULXs, in search of indications that may favour this newly emerging trend.
More specifically, we ask whether the curvature of ULX spectra is due to hot thermal emission, rather than a hard power law with a cutoff at an improbably low energy and whether this can be physically interpreted in terms of emission from a super-critically accreting NS.
To investigate this hypothesis, we have selected eighteen  well known ULXs that have been studied by multiple authors in the past and were also included in the samples used in the seminal works of \cite{2006MNRAS.368..397S} and \cite{2013MNRAS.435.1758S}, in which the spectral shape of ULXs was standardised and classified observationally. Below (Section~\ref{sec:origin}) we briefly present the different interpretations of the spectral curvature in ULXs, the latest theoretical arguments for the nature of their central engine and the motivation behind our choice to revisit the spectra of known ULXs. In Section~\ref{sec-observations} we present the details of our data analysis and in Section~\ref{discussion} we discuss our findings and their implications with regard to the nature of the accretor in our sample and in ULXs in general.

\section{Origin of the curvature in the spectra of ULXs}
\label{sec:origin}

\subsection{Optically thick, ``warm'' corona}

Nominal BH-XRBs exhibit a power-law shaped tail towards high energies, during all spectral states. More specifically, during episodes of low-luminosity-advection-dominated
accretion (also known as a {\it hard state} ) the spectrum is dominated by a hard power law (spectral index of up to ${\sim}$1.5) with an exponential cutoff at ${\sim}$100-200\,keV, while in the high-luminosity {\it soft state}, the power-law component becomes, less prominent and softer (spectral index exceeding ${\sim}$2), but without an observable spectral cutoff \citep[e.g.][]{2001AdSpR..28..307B,2003MNRAS.344...60G,2005Ap&SS.300..177N,2006ARA&A..44...49R, 2007A&ARv..15....1D,2010LNP...794...17G}. As discussed above, the spectra of many ULXs (including all sources studied in this work) feature a spectral curvature and an abrupt drop in spectral counts at considerably lower energies than standard BH-XRBs. The spectral roll-over in ULXs was detected in \xmm data of numerous sources \citep[e.g.][]{2006MNRAS.368..397S,2013MNRAS.435.1758S} and was subsequently confirmed with the \nus telescope \cite[e.g.][]{2013ApJ...778..163B,2014ApJ...793...21W,2015ApJ...806...65W,2017ApJ...834...77F}. It is often observed as low as ${\sim}$3\,keV.

 In the past, many authors have modelled the unusual low-energy curvature of ULX spectra (including the ones revisited here) using a power-law spectrum with a low-energy cutoff and a relatively large e-folding energy \citep[e.g.][]{2006MNRAS.368..397S,2007Ap..SS.311..203R,2009MNRAS.397.1836G,2013MNRAS.435.1758S}. The uncommonly low-energy cutoff, is often linked to the presence of a corona of hot, thermal electrons with an unusually high Thomson scattering optical depth. Namely,  multiple authors have considered that the shape of the high-energy part of the spectrum is the result of thermal Comptonisation of soft ($h\nu<0.5$\,keV) photons, by a corona of thermal electrons with an optical depth that often exceeds $\tau_{\rm T}\approx20$ \citep[e.g.][]{2006MNRAS.368..397S,2007Ap..SS.311..203R,2009MNRAS.397.1836G,2012MNRAS.420.1107P,2013MNRAS.435.1758S,2014MNRAS.439.3461P}. While this configuration successfully reproduces the observed spectral shapes, its feasibility under realistic circumstances in the vicinity of critically accreting X-ray binaries may be problematic.  More specifically, the high scattering depth combined with the increased photon density will pose a significant burden to the coronal thermalisation. This can be illustrated  in the following simplified example, where we consider cooling due to Inverse Compton (IC) scatterings and Coulomb collisions as the sole thermalisation mechanism. The cooling rate of thermal electrons due to multiple IC scatterings by photons with $h\nu\ll kT_{e}$ depends strongly on the scattering optical depth of the corona, that is,~the cooling timescale for large optical depth is \citep[e.g.][]{1979rpa..book.....R}
\begin{equation}
 t_{\rm cool}\approx \frac{m_{\rm e}c^{2}(R/c)}{4h\nu\,{\tau_{\rm T}}^{2}},
 \label{eq:tcool}
\end{equation}
where $m_{\rm e}$ is the electron mass and $R/c$ the characteristic source size, which can be inferred from variability considerations. It is obvious from Eq.~\ref{eq:tcool} alone that for the values of $\tau_{\rm T}$ reported in these works, Compton cooling will be very rapid, that is,~$t_{\rm cool}\ll{R/c}$. Below we illustrate that the cooling timescale will be too small to allow for electron thermalisation. 

The thermalisation will occur primarily through energy exchange between high-energy electrons and the thermal background of electrons and protons in the coronal plasma. For a high scattering optical depth (i.e.~${\tau}_{T}>5$), one can plausibly assume that the main mechanism for the energy exchange will be Coulomb interactions between electrons and electrons, and electrons and protons, ignoring, for example, collective interactions between particles \citep[e.g.][]{1988ApJ...332..872B}. If the relativistic electrons exchange energy more rapidly than they cool due to multiple scatterings, then the plasma will thermalise efficiently. The timescale of the Coulomb energy exchange will be
\begin{equation}
 t_{exch}{{\sim}}\frac{E}{\left |dE/dt  \right |},
 \label{eq:tcul}
\end{equation}
where E is the electron energy and dE/dt is the Coulomb cooling rate \citep{1975ApJ...196..689G,1979PhRvA..20.2120F,1990MNRAS.245..453C}. For relativistic electrons the Coulomb rate can be rewritten as \citep[see also][]{1999ASPC..161..375C}, 
\begin{equation}
\frac{dE}{dt}{{\sim}}-\tau_{\rm T}\ln{ \Lambda}{\left ( \frac{R}{c} \right )}^{-1}
 \label{eq:culrate}
,\end{equation}
where $\ln{ \Lambda}$ is the usual Coulomb logarithm. From the above approximations, it becomes evident that while the Coulomb cooling becomes more rapid, as optical depth increases ($t_{exch}{{\sim}}{\tau_{\rm T}}^{-1}$), the ${\tau_{\rm T}}^{-2}$ dependence of the Compton cooling timescale, results in a corona in which energetic photons cool down before they can thermalise. As ${\tau_{\rm T}}$ increases, a thermal corona becomes more and more difficult to sustain.
The problem of coronal thermalisation is well known, and it can also become significant in the low $\tau_{\rm T}$ regime \citep[e.g.][and references therein]{1999ASPC..161..375C}, particularly for hot ($kT_{e}{{\gtrsim}}{m_{\rm e}}c^{2}$) coronas. As a result, the presence of hybrid thermal/non-thermal electron distributions \citep[e.g.][]{1999ASPC..161..375C,2008AA...491..617B,2009MNRAS.392..570M} is usually assumed. 
However, the high-scattering optical depths that were claimed in earlier ULX literature  \citep[e.g.][]{2006MNRAS.368..397S,2007Ap..SS.311..203R,2009MNRAS.397.1836G,2013MNRAS.435.1758S}, present major sustainability issues even when considering hybrid, ``warm'' coronas with lower electron temperatures. It is only under very tight restrictions that coronas with $\tau_{\rm T}>5$ can be sustained \citep[e.g.][]{2015A..A...580A..77R}. 
We must stress here, that the issues concerning the physical plausibility of the optically thick corona model have been noted by the community since relatively early on \citep[e.g.][]{2011MNRAS.417..464M,2012MNRAS.422..990K}, with the majority of later works, only using the power law with the exponential cutoff as an empirical description of the hard emission of ULXs, rather than a physical interpretation.

\subsection{Optically thick winds}
 The most widely accepted interpretation of the spectral curvature of ULXs invokes the presence of strong, optically thick winds. Namely, under the assumption that ULXs are accreting BHs in the super-Eddington regime, then the spectral shape of the emission may be strongly influenced by the presence of
massive, optically thick outflows. \cite{2003MNRAS.345..657K} and \cite{2007MNRAS.377.1187P} argued that the curvature of the spectra (of at least some) ULXs can be interpreted in terms of reprocessing of the primary emission in the optically thick wind. In this scenario the soft thermal emission is associated with the wind itself, and the hard emission is also thermal and originates in the hot, innermost parts of an accretion disk. \cite{2009MNRAS.398.1450K} followed up the predictions of \cite{2007MNRAS.377.1187P} and by studying the spectra of eleven known ULXs they claimed that the temperature  of the soft thermal component decreased with its luminosity (i.e.~$L_{soft}{{\sim}}{T_{in}}^{-3.5}$), in agreement with the prediction for emission from an optically thick wind. However, subsequent studies by \cite[e.g.][]{2013ApJ...776L..36M} found that the luminosity of the soft component correlates positively with temperature, approximately following the $L_{soft}{{\sim}}{T_{in}}^{4}$ relations expected for standard accretion disks. However, in a recent study of numerous long-term observations of HO IX X-1, \cite{2016MNRAS.460.4417L} show that -- with increasing luminosity -- the source spectra evolve from a two-component spectrum to a (seemingly) single-component, thermal-like spectrum. The authors argue that the apparent heating of the soft-disk component may be model dependent, an artifact caused by  neglecting to properly account for the evolving spectra.

In addition to the prediction of the ULX spectral shape, the optically thick wind model also offers an interpretation for the short-timescale variability noted in many sources \citep[e.g.][]{2013MNRAS.435.1758S}. More specifically, \cite{2015MNRAS.447.3243M} combined the arguments of \cite{2003MNRAS.345..657K} and \cite{2007MNRAS.377.1187P} with predictions for wind inhomogeneity \citep[e.g.][]{2013PASJ...65...88T} and mass-accretion rate fluctuations \citep[e.g.][]{2013MNRAS.434.1476I} to  account for spectral and timing variability of nine ULXs. The authors made a well-founded case for super-Eddington accretion onto stellar-mass BHs as the driving force behind ULXs.  In this scheme the fractional variability noted by previous authors is explained in terms of a ``clumpy'' wind partially obscuring the hard component, which appears to fluctuate. In the same context the different empirical states indicated by \cite{2013MNRAS.435.1758S} are the result of different orientations between the observer and the disk/wind structure \citep[see Fig.~1 of][]{2015MNRAS.447.3243M}. Building on these considerations several authors \citep[e.g.][]{2014ApJ...793...21W,2015ApJ...806...65W,2016MNRAS.460.4417L} have modelled  ULX spectra extracted from \xmm and \nus data using a dual thermal model, in which the soft thermal emission is attributed to the optically thick wind and the hotter component to emission from the inner parts of a hot accretion disk.
The presence of the strong outflows are also supported by the uniform optical counterparts of numerous ULXs, which indicate a hot wind origin  \citep[e.g.][]{2002astro.ph..2488P,2015NatPh..11..551F}, but also the presence of wind or jet blown, radio ``bubbles'' around some ULXS \citep[e.g.][]{2003RMxAC..15..197P,2006MNRAS.368.1527S,2012ApJ...749...17C}.
The strongest indication of an outflow lies in the discovery of soft X-ray spectral residuals near ${\sim}1$\,keV \citep[e.g.][]{2014MNRAS.438L..51M,2015MNRAS.454.3134M,2016Natur.533...64P,2017MNRAS.468.2865P}, which have been interpreted as a direct signature of their presence. While broad emission- and absorption-like residuals near the ${\sim}1$\,keV mark have been observed in the spectra of numerous NS- and BH-XRBs during different states and at different mass accretion rates (e.g. \citealt{2003A&A...407.1079B}; \citealt{2006A&A...445..179D}; \citealt{2010A&A...522A..96N}; \citealt{2014MNRAS.437..316K}), the absorption lines detected in the spectra of NGC~1313 X-1 and NGC~5408 X-1 \citep{2016Natur.533...64P}, and more recently in NGC~55 ULX \citep{2017MNRAS.468.2865P}, seem to be strongly blue-shifted, indicating outflows with velocities reaching 0.2\,c.

\subsection{The case of accreting neutron stars}

The recent discoveries of the three PULXs, has established the fact that ULXs can be powered by accretion onto NSs. This realisation is perhaps not surprising, considering that a mechanism that allows accretion at super-Eddington rates onto highly magnetised (B${{\gtrsim}}10^{12}$G) NSs has been put forward since the late seventies \citep{1973A&A....25..233G,1975A&A....42..311B,1976MNRAS.175..395B}.
However, these works were not aimed at discussing super-Eddington accretion in the context of ULXs. The authors were attempting to resolve the complications that arise from the fact that when material is accreted onto a very small area on the surface of the NS the Eddington limit is considerably lower than the ${\sim}1.8\,10^{38}$\,erg/s, which corresponds to isotropic accretion onto a NS. Therefore, persistent X-ray pulsars with luminosities exceeding a\,few\,$10^{37}$\,erg/s, were in fact breaking the (local) Eddington limit.
More specifically, in high-B accreting NSs, the accretion disk is interrupted by the magnetic field near the magnetospheric radius, at which point the accreted material is guided by the magnetic field lines onto a small area centered around the NS magnetic poles \citep[e.g.][]{1972A&A....21....1P,2012MNRAS.421...63R}. The resulting formation is known as an accretion column. Due to the high anisotropy of the photon–electron scattering cross-section in the presence of a strong magnetic field \citep{1971PhRvD...3.2303C,1974ApJ...190..141L}, the emission from the accretion column is concentrated in a narrow (``pencil-'') beam, which is directed parallel to the magnetic field lines (and hence the magnetic field axis, \citealt{1975A&A....42..311B}). However, at high mass-accretion rates a radiation dominated shock is formed at a few km above the surface of the NS. As accretion rate exceeds a critical value \citep[corresponding to a critical luminosity of ${\sim}10^{37}$erg/sec, (e.g.][]{1976MNRAS.175..395B,2015MNRAS.447.1847M}, the accretion funnel is suffused with high-density plasma which is gradually sinking in the gravitational field of the NS \citep{1976MNRAS.175..395B,1981A&A....93..255W}. As a result, the accretion funnel, in the direction parallel to the magnetic field axis, becomes optically thick and the emerging X-ray photons mostly escape from its -- optically thin -- sides, in a ``fan-beam'' pattern \citep[see e.g. Fig.1][]{2007A&A...472..353S}. In recent refinements of this mechanism, it has been demonstrated that depending on the magnetic field strength and the pulsar spin it can facilitate luminosities of the order of $10^{40}$erg/sec \citep[][]{2015MNRAS.454.2539M}.

Observations of multiple X-ray pulsars have yielded an empirical description of the primary emission of the accretion column as a very hard power law (spectral index ${\lesssim}1.8$) with a low-energy (${\lesssim}10$\,keV) cutoff \citep[e.g.][]{2012MmSAI..83..230C}. While a general, self consistent description of the spectral shape of the accretion column emission has not been achieved yet, several authors have attempted to reproduce it \citep[e.g.][]{1981ApJ...251..288N,1985ApJ...299..138M,1991ApJ...367..575B,2004ApJ...614..881H,2007ApJ...654..435B}. More specifically, \cite{2007ApJ...654..435B} have reproduced the spectrum, assuming bulk and thermal Comptonisation of seed Bremsstrahlung, black body and cyclotron photons. 

While, in the high-field regime, super-Eddington accretion can be efficiently sustained, lowly magnetised NSs can only reach moderately super-Eddington luminosities and only in the soft state of the so-called Z-sources \citep[e.g.][]{2002ApJ...568L..35M,2007A&ARv..15....1D,2009ApJ...696.1257L}. 
When the accretion rate reaches and exceeds the Eddington limit, strong outflows are expected to inhibit higher accretion rates.
Nevertheless, in a recent publication, \cite{2016MNRAS.458L..10K} argue that super-Eddington accretion onto lowly magnetised NSs can be maintained -- along with powerful outflows -- in a similar fashion to super-Eddington accretion onto BHs \citep{2001ApJ...552L.109K}. Therefore, a considerable fraction of (non-pulsating) ULXs may be the result of beamed emission from NSs with relatively weak magnetic fields (B$<10^{11}$\,G), accreting at high mass-transfer  rates. 

Intriguingly, one of the first \citep{1984PASJ...36..741M} and most frequently observed spectral characteristics associated with the soft state of Z-sources is the presence of two thermal components \citep[e.g.][]{1988ApJ...324..363W,1989PASJ...41...97M,2001AdSpR..28..307B,2007ApJ...667.1073L,2013MNRAS.434.2355R}. The ``cool'' thermal component most likely originates in a thin \citeauthor{1973A&A....24..337S} disk and the additional ``hot'' thermal component corresponds to emission produced in hot optically thick plasma on the the surface of the NS, known as a {\it boundary layer} \citep{1986SvAL...12..117S,2000AstL...26..699S}. 
Therefore, the presence of a dual thermal spectrum in XRBs strongly suggests emission from a solid surface, indicating a lowly magnetised NS. Nevertheless, in a new publication by \cite{2017MNRAS.tmp..143M} it is argued  that in accreting high-B NSs, the normally optically thin \citep[e.g.][]{1980A&A....87..330B,1992ApJ...396..147N,1996PASJ...48..425E}  material trapped in the Alfv{\'e}n surface becomes optically thick as the luminosity exceeds ${{\sim}}5\,10^{39}$\,erg/s. The emission of the resulting structure will have a quasi-thermal spectrum at temperatures exceeding 1\,keV. 
Combined with the soft thermal emission from a truncated accretion disk, the spectra of highly magnetised NSs may also feature the same dual thermal shape as high-state Z-sources \citep[][see more details in Sections~\ref{sec-observations} and~\ref{discussion}]{2017MNRAS.tmp..143M}. 
Based on these considerations, it becomes apparent that the reanalysis of ULX spectra is warranted. To this end, we explore the relevance of models, usually implemented in the modelling of emission from NS-XRBs, in the context of ULXs. More importantly, we investigate whether or not our best-fit models yield parameter values that are physically meaningful and in accordance with the predictions for the emission from ULXs.

\section{Spectral extraction and data analysis}
\label{sec-observations}
We have analysed  archival \xmm observations of eighteen ULXs presented in Table~\ref{tab:log}.  We have selected sources that have been studied extensively in the past and are confirmed ULXs.  Furthermore, specific datasets were selected to have a sufficient number of counts to allow robust discrimination between the different models used to describe their spectral continuum. Except for these conditions, sources were chosen randomly.  Therefore, the ULX sample analysed in this work is not complete. Nevertheless, the purpose of this work is not to revisit the entire ULX catalogue, but to demonstrate that a significant fraction of ULXs follow a specific pattern (presented below). For this purpose, our source sample is sufficiently extensive. For six of these sources  we also analysed archival \nus data.

\subsection{\xmm spectral extraction}
For the \xmm data, we only considered the EPIC-pn detector, which has the largest effective area of the three EPIC detectors, in the full 0.3-10\,keV bandpass, and registered more than ${\sim}$1000 photons for each of the observations considered, thus providing sufficient statistics to robustly discriminate between different spectral models while ensuring simplicity and self-consistency in our analysis.
Therefore the following description of data analysis refers only to this instrument. The data were handled using the \xmm data analysis software SAS version 15.0.0. and the calibration files released\footnote{XMM-Newton CCF Release Note: XMM-CCF-REL-332}  on January 22, 2016. In accordance with the standard procedure, we filtered all observations for high background-flaring activity, by extracting high-energy light curves (10$<$E$<$12\,keV) with a 100\,s bin size. By placing appropriate threshold count rates for the high-energy photons, we filtered out time intervals that were affected by high particle background.
During all observations pn was operated in Imaging Mode. In the majority of our sources, spectra were extracted from a circular region with a radius $>$35\arcsec\  centred at the core of the point spread function (psf) of each source. We thus ensured the maximum encircled energy fraction\footnote{See \xmm Users Handbook \S3.2.1.1 \\  http://xmm-tools.cosmos.esa.int/external/xmm\_user\_support/ \\ documentation/uhb/onaxisxraypsf.html} within the extraction region. This was not possible in the case of NGC~4861 ULX1, M81~X-6, and NGC~253 ULX2 where we used spectral extraction apertures of $18{\arcsec}$, 13.8\arcsec and  12.5\arcsec in order to avoid contamination by adjacent sources or due to the proximity of our source  to the edge of the chip\footnote{When part of the psf lies in a chip gap, effective exposure and encircled energy fraction may be affected}. The extraction and filtering process followed the standard instructions provided by the \xmm Science Operations Centre (SOC\footnote{http://www.cosmos.esa.int/web/xmm-newton/sas-threads}). More specifically, spectral extraction was done with SAS task \texttt{evselect} using the standard filtering flags (\texttt{\#XMMEA\_EP \&\& PATTERN<=4} for pn), and SAS tasks \texttt{rmfgen} and \texttt{arfgen} were used to create the redistribution matrix and ancillary file,  respectively. All spectra were regrouped to have at least 25 counts per bin and analysis was performed using the {\tt xspec} spectral fitting package, version 12.9.0 \citep{1996ASPC..101...17A}.

\begin{table*}[!htbp]
 \caption{Observation Log.}
 \begin{center}
\scalebox{0.8}{   \begin{tabular}{lcccccccc}
     \hline\hline\noalign{\smallskip}
     \multicolumn{1}{c}{Source} &
     \multicolumn{1}{c}{Distance$^{a}$} &
     \multicolumn{1}{c}{ObsID} &
     \multicolumn{1}{c}{Date} &
     \multicolumn{1}{c}{Duration$^{b}$} &
     \multicolumn{1}{c}{Rate$^{c}$} &
     \multicolumn{1}{c}{Obs. Mode} &
     \multicolumn{1}{c}{Filter} &
     \multicolumn{1}{c}{Position}\\
     \noalign{\smallskip}\hline\noalign{\smallskip}
      
      \multicolumn{1}{c}{} &
      \multicolumn{1}{c}{Mpc} &
      \multicolumn{1}{c}{} &
      \multicolumn{1}{c}{} &     
      \multicolumn{1}{c}{s} &
      \multicolumn{1}{c}{${\rm 10^{-1}\,{ct}\,{s}^{-1}}$} &
      \multicolumn{1}{c}{} &
      \multicolumn{1}{c}{} &
      \multicolumn{1}{c}{} \\
      \noalign{\smallskip}\hline\noalign{\smallskip}

      \noalign{\smallskip}\hline\noalign{\smallskip}
      \xmm       & &   &  &   &   & & &  \\
      \noalign{\smallskip}\hline\noalign{\smallskip}
       Ho II X-1        &3.27$\pm 0.60$ &  0200470101   & 2004-04-15 & 44130  & 6.97$\pm 0.04$  & FF  & Medium & On-Axis \\
       Ho IX X-1        &3.77$\pm 0.80$ &  0200980101   & 2004-09-26 & 75900  & 14.5$\pm 0.05$  & LW  & Thin   & On-Axis    \\
       IC~342 X-1       &2.73$\pm 0.70$ &  0206890201   & 2004-08-17 & 16970  & 3.96$\pm 0.05$  & EFF & Medium & On-Axis    \\
       M33 X-8          &0.91$\pm 0.50$ &  0102640101   & 2000-08-04 & 7144   & 55.1$\pm 0.28$  & FF  & Medium & On-Axis    \\
       M81 X-6          &3.61$\pm 0.50$ &  0111800101   & 2001-04-22 & 79660  & 4.38$\pm 0.02$  & SW  & Medium & On-Axis     \\
       M83 ULX          &4.66$\pm 0.70$ &  0110910201   & 2003-01-27 & 19760  & 1.20$\pm 0.03$  & EFF & Thin   & ${\sim} 6.5\arcmin$    \\
       NGC~55 ULX       &1.60$\pm 0.20$ &  0028740201   & 2001-11-14 & 34442  & 1.25$\pm 0.07$  & FF  & Thin   & On-Axis \\
       NGC~253 ULX2     &3.56$\pm 0.80$ &  0152020101   & 2003-06-19 & 65010  & 2.57$\pm 0.02$  & FF  & Thin   & On-Axis    \\
       NGC~253 XMM2     &3.56$\pm 0.80$ &  0152020101   & 2003-06-19 & 64850  & 2.41$\pm 0.02$  & FF  & Thin   & ${\sim} 4.1\arcmin$   \\
       NGC~1313 X-1     &4.25$\pm 0.80$ &  0106860101   & 2000-10-17 & 19880  & 6.89$\pm 0.01$  & FF  & Medium & ${\sim} 5.4\arcmin$   \\
       NGC~1313 X-2     &4.25$\pm 0.80$ &  0405090101   & 2006-10-15 & 80470  & 6.19$\pm 0.03$  & FF  & Medium & ${\sim} 4.0\arcmin$   \\
       NGC~4190 ULX1    &2.83$\pm 0.10$ &  0654650301   & 2010-11-25 & 11070  & 12.6$\pm 0.11$  & FF  & Medium & On-Axis    \\
       NGC~4559 X-1     &7.31$\pm 0.20$ &  0152170501   & 2003-05-27 & 33950  & 2.72$\pm 0.03$  & FF  & Medium & On-Axis   \\
       NGC~4736 ULX1    &4.59$\pm 0.80$ &  0404980101   & 2006-11-27 & 35040  & 1.89$\pm 0.02$  & FF  & Thin   & On-Axis   \\
       NGC~4861 ULX     &7.00$\pm 1.00$&  0141150101   & 2003-06-14 & 13180  & 0.73$\pm 0.02$  & FF  & Medium & On-Axis  \\
       NGC~5204 X-1     &4.76$\pm 0.90$ &  0405690101   & 2006-11-15 & 7820   & 9.67$\pm 0.11$  & FF  & Medium & On-Axis \\
       NGC~5907 ULX     &17.1$\pm 0.90$&  0729561301   & 2014-07-09 & 37480  & 3.30$\pm 0.03$  & FF  & Thin   & On-Axis    \\
       NGC~7793 P13     &3.58$\pm 0.70.$&  0748390901   & 2014-12-10 & 41970  & 4.85$\pm 0.04$  & FF  & Thin   & ${\sim} 4.0\arcmin$     \\
      \noalign{\smallskip}\hline\noalign{\smallskip}
      {\it NuSTAR}      & &   &  &   &   & & &  \\
      \noalign{\smallskip}\hline\noalign{\smallskip}
       Ho II X-1        &3.27$\pm 0.60$ &  30001031005  & 2013-09-17 & 111104 & 0.38$\pm 0.01$  & --  & None   & ${\sim} 3.1\arcmin$  \\
       Ho IX X-1        &3.77$\pm 0.80$ &  30002033003  & 2012-10-26 & 88030  & 1.29$\pm 0.13$  & --  & None   & ${\sim} 1.6\arcmin$  \\
       IC~342 X-1       &2.73$\pm 0.70$ &  90201039002  & 2016-10-16 & 49173  & 1.26$\pm 0.02$  & --  & None   & On-Axis         \\
       NGC~1313 X-1     &4.25$\pm 0.80$ &  30002035002  & 2012-12-16 & 100864 & 6.19$\pm 0.03$  & --  & None   & ${\sim} 2.8\arcmin$        \\
       NGC~1313 X-2     &4.25$\pm 0.80$ &  30002035002  & 2012-12-16 & 100864 & 6.19$\pm 0.03$  & --  & None   & On-Axis        \\
       NGC~5907 ULX     &17.1$\pm 0.90$&  80001042002  & 2014-07-09 & 57113  & 0.26$\pm 0.01$  & --  & None   & On-Axis  \\    
      \noalign{\smallskip}\hline\noalign{\smallskip}                                                                                                                                                                                                         
      \noalign{\smallskip}\hline\noalign{\smallskip}         
      \noalign{\smallskip}\hline\noalign{\smallskip}

      \noalign{\smallskip}\hline\noalign{\smallskip}         
          
      \noalign{\smallskip}\hline\noalign{\smallskip}
    \end{tabular}   }
 \end{center}
  \tablefoot{
  \tablefoottext{a}{All distance estimations are from \cite{2013AJ....146...86T}}
  \tablefoottext{b}{Of filtered pn observation}.
  \tablefoottext{c}{Corresponding to the pn spectrum of the source}.}
 \label{tab:log}
\end{table*}

\subsection{{\nus} spectral extraction}

The \nus data were processed using version 1.6.0 of the \nus data analysis system (\nus DAS). 
We downloaded all public \nus datasets using the ``heasarc\_pipeline'' scripts (Multimission Archive Team{\@}OAC, in prep.). 
These have already been processed to obtain L1 products. We then ran {\tt nuproducts} using a 50\arcsec extraction region around  the main source and a 50-80\arcsec extraction region for background, in the same detector as the source when possible, as far as we could to avoid contributions from the point-spread function (PSF) wings. We applied standard PSF, alignment, and vignetting corrections. Spectra were rebinned in order to have at least 30 counts per bin to ensure the applicability of the $\chi^2$ statistics. 
All sources in our sample dominate the background up to 20-30\,keV. The models we use are relatively simple, and the physical interpretation does not change considerably for a change of best-fit parameters of 10-20\%, and so we do not need an extremely precise modelling of the background.

\subsection{Spectral analysis}

\subsubsection{\xmm}

The spectral continuum was modelled twice, firstly using a combination of a multicolour disk black body (MCD) and a black body component ({\tt diskbb+bbody}), with the black body ({\tt bbody}) acting as the hot thermal component and secondly using two MCD components ({\tt diskbb+diskbb}). The first model was used because it is the most widely used model describing the spectra of NS-XRBs in the high-accretion, ``soft'' state. Our choice for the second model was based on the recent theoretical predictions by \cite{2017MNRAS.tmp..143M}, where it is argued that critically accreting NSs with a high magnetic field (i.e.~B${\gtrsim}10^{12}$\,G) can become engulfed in an optically thick toroidal envelope which is the result of accreting matter moving along the magnetic field lines. Emission from the optically thick envelope is predicted to have a multicolour black body spectrum, with a temperature exceeding ${\sim}$1\,keV (more details in section~\ref{discussion}). We model this hot thermal emission using the {\tt diskbb} model because it is the simplest and most reliable multi-temperature black body model in {\tt xspec}; however we stress that we do not expect this emission to originate from a disk. Therefore, the inner disk radius corresponding to the hot {\tt diskbb} component has no physical meaning and is not tabulated. In both models the cool disk component is modelled as a geometrically thin, optically thick \cite{1973A&A....24..337S} disk, which is expected to extend inwards until it reaches the surface of the NS, unless it is disrupted by strong outflows or a strong magnetic field (more details in section~\ref{discussion}). We did not model intrinsic and/or host galaxy absorption separately from the Galactic absorption, but used one component for the total interstellar absorption, which was modelled using {\tt tbnew\_gas}, the latest improved version of the  {\tt tbabs} X-ray absorption model \citep{2011_in_prep}.

For the dual MCD model, we assume that the disk becomes truncated at  approximately the magnetospheric radius at which point the material follows the magnetic field lines to form the hypothetical envelope. Under this assumption, we also estimate the strength of the magnetic field (B), assuming that the inner radius of the ``cool'' {\tt diskbb} coincides with the magnetospheric radius ($R_{\rm mag}$) and using the expression for  $R_{\rm mag}$ given by \citeauthor{2014EPJWC..6401001L} (2014; see also Eq. 1 from \citealt{2017MNRAS.tmp..143M}). 
The complete {\tt xspec} model used in the spectral fits is {\tt tbnew\_gas(cflux*diskbb + cflux*(diskbb or bbody) )}, where {\tt cflux } is a convolution model that is used to calculate the flux of the two thermal components. Some of the sources exhibited strong residuals in the 0.5-1.2\,keV region, commensurate with X-ray emission lines from hot, optically thin plasma. The emission features were modelled using the {\tt mekal} model, which models the emission spectrum of a hot diffuse gas. The best fit parameters of the continuum were not sensitive to the modelling of these features (e.g. using a Gaussian instead of {\tt mekal}), however they are strongly  required by the fit (${\delta}{\chi}^2>15$ for two d.o.f in all sources).  More specifically, plasma temperature  was ${\sim}$0.92\,keV for Ho IX X-1,  ${\sim}$1.09\,keV for M81 X-6, ${\sim}$0.95\,keV for M83 ULX , ${\sim}$0.40keV for NGC~4736 ULX, and  ${\sim}$1.08\,keV for NGC~5204 X-1. While soft X-ray atomic features may be crucial to our understanding of the nature of ULXs \citep[namely, the presence of strong winds and the chemical composition of the accreted material, e.g.:][]{2015MNRAS.447.3243M, 2015MNRAS.454.3134M, 2016Natur.533...64P}, they are not the focus of this work and are only briefly discussed in Section~\ref{discussion}, but not studied further. 

Given the high (${{\gtrsim}}1$\,keV) temperatures of the hot thermal component in both double-thermal models, it is expected that electron scattering will have a significant effect on the resulting spectrum, as it becomes comparable to free-free absorption. Therefore, the actual emission will be radiated as a ``diluted'' black body, which when modelled using a prototypical thermal model like {\tt diskbb} or {\tt bbody}  will result in temperature and radius estimations that deviate from their ``true'' values. This issue is commonly addressed by considering a correction factor (\f) that approximately accounts for the spectral modification \citep{1986ApJ...306..170L,1986SvAL...12..383L,1995ApJ...445..780S}; this factor is often referred to as a {\it colour correction factor} and detailed calculations, combined with multiple observations have demonstrated that it depends weakly on the size\footnote{or inner radius in the case of an accretion disk} of the emitting region and the mass accretion rate \citep[e.g.][]{1995ApJ...445..780S}. Therefore in the first approximation it can be considered independent of these parameters and its value is estimated between ${\sim}1.5$ and ${\sim}$2.1 \citep[e.g.][and references therein]{2005ApJ...618..832Z}. The colour correction factor affects both the temperature and normalisation of the thermal models (i.e.~{\tt diskbb} and {\tt bbody}), with the corrected values given by
\begin{equation}
T_{\rm cor}=\frac{T}{f_{col}}
\label{eq:Tcor}
,\end{equation}
and
\begin{equation}
R_{\rm in,cor}=R_{\rm in}\,{f_{\rm col}}^2
\label{eq:Rcor}
,\end{equation}
where $T$ is the temperature of the MCD component and $R_{\rm in}$ is the inner radius. Although the spectral hardening effects are expected to be noticeable, particularly in the hot thermal component, we have decided not to include the colour correction in any of our calculations and to tabulate and plot the values of all quantities of interest as provided by our best fits. The reader is, however, advised to note that the value of our results may vary by a value of ${\sim}${\f}.

The value of the inferred inner disk radius is also dependent on the viewing angle ($i$) of the accretion disk (i.e.~$R_{in}{{\sim}}1/\sqrt{\cos{i}}$). This dependence may become important in the estimation of ${\rm R_{in}}$ if the accretion disk is viewed at a large inclination angle (i.e.~edge-on view).  Nevertheless, since we have no indications for a high viewing angle  (e.g. dips\footnote{With the exception of NGC~55 ULX, which does show dips in its light curve \citep{2004MNRAS.351.1063S}. However, due to its likely supercritical accretion rates, the dips are not as constraining, for its inclination, as in typical XRBs.} in their light curves or spectral absorption features resulting from an edge-on view of the accretion disk atmosphere) in any of our sources, we have selected a value of $i=60\deg$ for all sources in our list.

All best fit values for the absorbed dual-MCD model together with the estimations for the magnetic field and their classification, as proposed by \cite{2013MNRAS.435.1758S}, are presented in Table~\ref{tab:xmm_fit}. The values for the absorbed MCD/black body model are presented  in Table~\ref{tab:bbody}. In Table~\ref{tab:bbody} we also provide the estimations of the ``spherization'' radius \citep{1973A&A....24..337S} of each source. Lastly, we note that the $\chi^2$ values from the {\tt tbnew\_gas*(diskbb+bbody)} fits were similar, albeit moderately higher than those of the dual MCD model and with moderately lower temperature of the hot component ($kT_{\rm BB}$ between ${\sim}$0.9\,keV and ${\sim}$2.2\,keV).

\subsubsection{{\it NuSTAR}}

In the {\it NuSTAR} spectra we ignored all channels below 4\,keV, and thus we did not require the addition of the cool thermal component. The primary spectral component used to model all spectra is again a multicolour disk black body ({\it diskbb}), the ``hot'' thermal component from the \xmm fits.  Furthermore,  we also look for the presence of a potential hard, non-thermal tail, which is usually detected in most XRBs, even in the soft state. A simultaneous  broadband fit of the combined {\it NuSTAR} plus \xmm spectra is not explored in this paper. Recent, rigorous works have extensively studied the \xmm (or \swift) + \nus data that we revisit here (Ho~II~X-1: \citealt{2015ApJ...806...65W}; HoIX~X-1: \citealt{2016MNRAS.460.4417L}; IC~342~X-1: \citealt{2015ApJ...799..121R}; NGC~1313~X-1,X-2: \citealt{2013ApJ...778..163B}; NGC~5907~X-1: \citealt{2017ApJ...834...77F}) and have  noted the presence of a spectral shape that can be modelled either as a hot thermal component or sharp cutoff with an additional, weak power-law tail. In this paper we do not seek to reproduce these analyses, but to discuss a possible novel interpretation of the spectral shape. The {\it NuSTAR} data  are used with the purpose of confirming (or dismissing) the presence of these components in a comprehensive and consistent study. To this end, the separate analysis is swift and effective.

\begin{figure}
       \resizebox{\hsize}{!}{\includegraphics[angle=-90,clip,trim=0 0 0 0,width=0.8\textwidth]{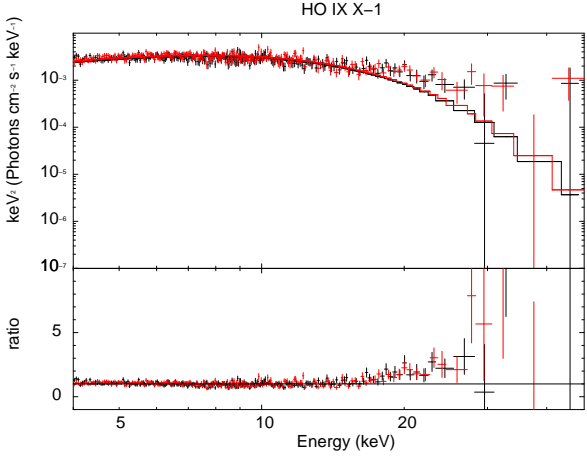}}
            \caption{Ho IX X-1: Unfolded spectrum. Energy and data-vs-model ratio plot, for only the {\tt diskbb} model. There are clear residuals above 20\,keV. }
   \label{fig:res}
 \end{figure}

\begin{figure}
       \resizebox{\hsize}{!}{\includegraphics[angle=-90,clip,trim=0 0 0 0,width=0.8\textwidth]{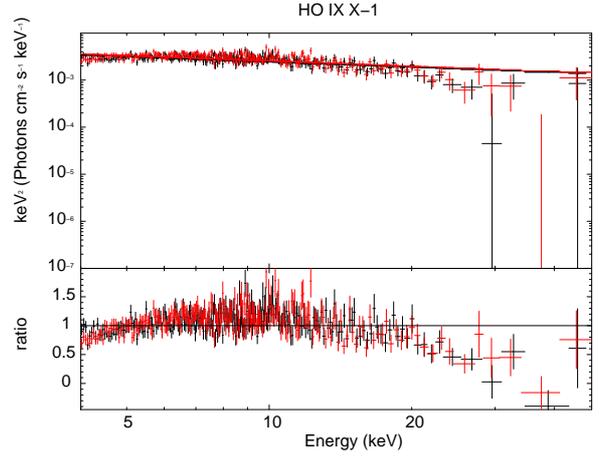}}
            \caption{Ho IX X-1: Unfolded spectrum. Energy and data-vs-model ratio plot, for only the power law model. There is a clear curvature in the spectrum that cannot be described by the power law. }
   \label{fig:po}
 \end{figure}

The \nus data were modelled using a single {\it diskbb} model and an additional power law. The hard ($>$10\,keV) power-law emission is faint, with less than 5\% of observed photons registered above 20\,keV, on average. Furthermore, the background contamination becomes predominant above 25\,keV. Therefore, the slope or even the exact shape (i.e.~the presence of an exponential cutoff) of the hard spectral tail cannot be constrained accurately. Both the thermal component and the power-law tail are required, in order to achieve an acceptable fit. More specifically, fitting the \nus data with only the {\tt diskbb} model results in pronounced residuals above ${{\sim}}15$\,keV (e.g. Figure~\ref{fig:res}) and a value for the reduced $\chi^2$ that exceeds ${\sim}$1.3 in all sources. Similarly, fitting the \nus spectra with just a power-law, results in residual structure characteristic of thermal emission (Figure~\ref{fig:po}) and reduced $\chi^2$ values exceeding ${\sim}$1.2. 

The temperature gradient of the accretion curtain will, most likely, differ from the $T{{\sim}}r^{-0.75}$ predicted by the standard thin disk MCD models like {\tt diskbb} and this deviation will be enhanced further by electron scatterings. To diagnose the impact of this effect on the registered spectra, we also fitted them with the {\tt xspec} model {\tt diskpbb}, in which the disk temperature is proportional to $r^{-p}$ and p is a free parameter. We find that the value of $p$ is significantly smaller (on average $p{{\lesssim}}0.53$) than that of a standard thin accretion disk. More interestingly, we find that the {\tt diskpbb} fits did not require the addition of a power-law tail and yielded the same $\chi^2$ values as the {\tt diskbb + powerlaw} fits; albeit with the notable exception of NGC~5907, which is the only pulsating ULX in our \nus sample. In principle, the {\tt diskpbb} model could also be used to model the cool thermal emission, detected in the \xmm data, since the inner disk parts may also become inflated due to the high accretion rates (see discussion in Sect.~\ref{discussion}). Nevertheless, the addition of an extra free parameter in each thermal component will only add to the degeneracy between different models and will not provide any further insight into the physical parameters (i.e.~temperature and size) of the emitting regions. Therefore, the {\tt diskpbb} model is only used as a diagnostic for the geometry of the accretion curtain, and only for the \nus data where its impact is much more significant; its implications are discussed further in Section~\ref{discussion}. While we have analysed all available {\it NuSTAR} data for our sources, we only tabulate the results for those observations with the largest number of counts and for luminosities closer to the \xmm observations. Nevertheless all {\it NuSTAR} observations produced -- more or less -- similar results to the ones presented here (see observation log in Table~\ref{tab:log}, for the {\it NuSTAR} observations analysed in this work). Best fit values (including the value of $p$ and the temperature of the {\tt diskpbb} models) for the {\it NuSTAR} data are presented in Table~\ref{tab:nus_fit}.

\begin{table*}[!htbp]
 \caption{Best fit parameters of the dual MCD model for the \xmm observations. All errors are in the 1$\sigma$ confidence range.}
 \begin{center}
\scalebox{0.8}{   \begin{tabular}{lccccccccccc}
     \hline\hline\noalign{\smallskip}
     \multicolumn{1}{c}{Source} &
     \multicolumn{1}{c}{nH} &
     \multicolumn{1}{c}{k${\rm T_{disk}}$} &
     \multicolumn{1}{c}{${\rm R_{disk}}^{a}$} &
     \multicolumn{1}{c}{k${\rm T_{hot}}$} &
     \multicolumn{1}{c}{${\rm {L/L_{edd}}^{b}}$} &
     \multicolumn{1}{c}{${\rm L_{\rm hot}/L_{\rm disk}}$} &   
     \multicolumn{1}{c}{$B$} &  
     \multicolumn{1}{c}{Clas.$^{c}$} &  
     \multicolumn{1}{r}{red. $\chi^2/dof$ }\\
     \noalign{\smallskip}\hline\noalign{\smallskip}
      
      \multicolumn{1}{c}{} &
      \multicolumn{1}{c}{[$\times10^{21}$\,cm$^2$]} &     
      \multicolumn{1}{c}{keV} &
      \multicolumn{1}{c}{[km]} &
      \multicolumn{1}{c}{keV} &
      \multicolumn{1}{c}{} &
      \multicolumn{1}{c}{} &
      \multicolumn{1}{c}{$10^{12}$G} &  
      \multicolumn{1}{c}{} &   
      \multicolumn{1}{c}{} \\
      \noalign{\smallskip}\hline\noalign{\smallskip}

      \noalign{\smallskip}\hline\noalign{\smallskip}

       Ho~II X-1        &   0.40$^{d}$              & 0.42$_{-0.01}^{+0.02}$ & 1421$_{-112 }^{+128 }$ & 1.65$_{-0.05}^{+0.06}$ &   69.8$_{-17.7}^{+21.3}$ & 0.74$\pm0.03$ &   22.7$_{-2.98}^{+3.27}$&   SUL  & 1.22/137   \\
       Ho~IX X-1        &   0.64$_{-0.13}^{+0.15}$  & 0.48$_{-0.04}^{+0.05}$ &  557$_{-108 }^{+134 }$ & 3.15$_{-0.09}^{+0.10}$ &   78.5$_{-22.2}^{+27.5 }$ & 5.62$\pm0.32$&   4.64$_{-1.42}^{+2.09}$&   HUL  & 1.03/156   \\
       IC~342 X-1       &   7.89$_{-0.53}^{+0.60}$  & 0.48$_{-0.05}^{+0.06}$ &  431$_{-97.8}^{+132 }$ & 2.84$_{-0.09}^{+0.10}$ &   21.8$_{-7.45}^{+9.70 }$ & 2.38$\pm0.17$&   1.56$_{-0.55}^{+0.98}$&   HUL  & 1.02/102   \\
       M33 X-8          &   0.95$\pm0.01$           & 0.50$_{-0.07}^{+0.08}$ &  237$_{-47.8}^{+70.4}$ & 1.35$_{-0.05}^{+0.06}$ &   11.1$_{-7.73}^{+8.93}$ & 4.22$\pm1.11$ &   0.42$_{-0.13}^{+0.25}$&   BD & 0.99/121   \\
       M81 X-6          &   1.63$_{-0.15}^{+0.13}$  & 0.93$_{-0.30}^{+0.23}$ &  145$_{-19.2}^{+18.6}$ & 1.72$_{-0.27}^{+0.16}$ &   24.8$_{-5.74}^{+6.55}$ & 1.27$\pm0.81$ &   0.27$_{-0.06}^{+0.14}$&   BD & 1.13/124   \\
       M83 ULX          &   0.11$_{-0.10}^{+0.93}$  & 0.37$_{-0.12}^{+0.11}$ &  411$_{-76.1}^{+62.3}$ & 1.23$_{-0.13}^{+0.16}$ &   6.25$_{-1.39}^{+1.61}$ & 3.25$\pm2.18$ &   0.80$_{-0.37}^{+0.36}$&   --   & 0.92/44   \\
       NGC~55 ULX       &   2.40$_{-0.60}^{+0.59}$  & 0.28$\pm 0.02$         & 1163$_{-161}^{+108}$   & 0.82$\pm 0.02$         &   10.1$_{-3.22}^{+4.01}$ & 1.12$\pm0.10$ &   6.06$_{-2.81}^{+3.04}$&   SUL & 1.08/89  \\
       NGC~253 ULX2     &   3.59$_{-0.60}^{+0.59}$  & 0.24$_{-0.03}^{+0.05}$ & 1263$_{-594 }^{+661 }$ & 1.64$\pm 0.04$         &   18.5$_{-6.08}^{+7.56}$ & 8.99$\pm3.72$ &   9.89$_{-6.64}^{+8.54}$&   BD & 0.99/120  \\
       NGC~253 XMM2     &   1.17$_{-0.17}^{+0.19}$  & 0.53$_{-0.07}^{+0.08}$ &  239$_{-45.8}^{+65.4}$ & 1.54$_{-0.10}^{+0.14}$ &   10.2$_{-3.51}^{+4.40}$ & 2.60$\pm0.64$ &   0.41$_{-0.13}^{+0.21}$&   BD & 1.13/108   \\
       NGC~1313 X-1     &   2.42$_{-0.23}^{+0.24}$  & 0.32$\pm 0.02$         & 1612$_{-236 }^{+282 }$ & 2.33$_{-0.10}^{+0.11}$ &   52.3$_{-13.3}^{+16.1}$ & 2.41$\pm0.17$ &   24.3$_{-5.91}^{+7.91}$&   SUL  & 0.88/113   \\
       NGC~1313 X-2     &   1.83$_{-0.10}^{+0.11}$  & 0.65$_{-0.06}^{+0.07}$ &  313$_{-41.4}^{+52.3}$ & 2.25$_{-0.12}^{+0.16}$ &   50.2$_{-14.6}^{+17.5 }$& 3.23$\pm0.81$ &   1.44$_{-0.32}^{+0.45}$&   BD & 0.95/144   \\
       NGC~4190 ULX1    &   0.99$_{-0.47}^{+0.58}$  & 0.50$_{-0.12}^{+0.22}$ &  352$_{-162 }^{+299 }$ & 1.95$_{-0.10}^{+0.20}$ &   36.2$_{-2.16}^{+2.28}$ & 6.88$\pm3.02$ &   1.52$_{-0.96}^{+2.78}$&   BD & 1.00/112   \\
       NGC~4559 X-1     &   0.80$_{-0.31}^{+0.33}$  & 0.26$_{-0.02}^{+0.03}$ & 2609$_{-604 }^{+802 }$ & 1.41$_{-0.07}^{+0.08}$ &   42.5$_{-1.63}^{+1.59}$ & 1.99$\pm0.73$ &   49.1$_{-18.2}^{+29.3}$&   SUL  & 1.05/84   \\
       NGC~4736 ULX1    &   3.54$_{-1.62}^{+1.03}$  & 0.40$_{-0.09}^{+0.08}$ &  513$_{-178 }^{+373 }$ & 1.18$_{-0.08}^{+0.13}$ &   19.5$_{-4.46}^{+5.30}$ & 2.34$\pm0.51$ &   1.96$_{-0.73}^{+1.23}$&   BD & 1.09/68   \\
       NGC~4861 ULX     &   1.24$_{-0.90}^{+1.07}$  & 0.25$_{-0.05}^{+0.09}$ & 1639$_{-360 }^{+512 }$ & 1.13$_{-0.20}^{+0.44}$ &   12.6$_{-2.13}^{+2.51}$ & 1.52$\pm0.47$ &   11.4$_{-4.00}^{+6.96}$&   --   & 1.09/22  \\
       NGC~5204 X-1     &   0.18$_{-0.18}^{+0.41}$  & 0.36$_{-0.05}^{+0.04}$ & 1428$_{-322 }^{+567 }$ & 1.41$_{-0.12}^{+0.08}$ &   48.8$_{-12.5}^{+15.1}$ & 1.22$\pm0.16$ &   19.0$_{-6.85}^{+11.6}$&   SUL  & 1.09/64  \\
       NGC~5907 ULX     &   0.75$_{-0.09}^{+0.10}$  & 0.40$_{-0.06}^{+0.08}$ & 2497$_{-324 }^{+257 }$ & 2.53$_{-0.11}^{+0.13}$ &    483$_{-40.9}^{+39.7}$ & 5.51$\pm0.66$ &   184 $_{-12.9}^{+11.3}$&   --  & 1.09/83  \\
       NGC~7793 P13     &   1.25$_{-0.26}^{+0.27}$  & 0.33$_{-0.03}^{+0.04}$ &  671$_{-160 }^{+203 }$ & 3.27$_{-0.12}^{+0.13}$ &   34.6$_{-8.78}^{+10.6}$ & 11.0$\pm0.35$ &   4.17$_{-1.58}^{+2.46}$&   --   & 1.08/141  \\
      \noalign{\smallskip}\hline\noalign{\smallskip}         
      \noalign{\smallskip}\hline\noalign{\smallskip}

      \noalign{\smallskip}\hline\noalign{\smallskip}         
          
      \noalign{\smallskip}\hline\noalign{\smallskip}
    \end{tabular}   }
 \end{center}
  \tablefoot{
  \tablefoottext{a}{Radius inferred from cool {\texttt {diskbb}} component by solving K=${\rm(R_{\rm disk}/{D_{10kpc}})^{2}}\,\cos{i}$, for $R_{\rm disk}$ (the inner radius of the disk in km).  `K' is the normalisation of the cool \texttt{diskbb} model, ${\rm D_{10kpc}}$ is distance in units of 10\,kpc and `i' is the inclination.  }\\
  \tablefoottext{b}{``Bolometric'' luminosity (L) in the 0.1-100\,keV range, extrapolated from the best-fit model. ${\rm L_{edd}}$ is the Eddington luminosity for an isotropically accreting NS with a mass of $1.4\,{\rm M_{\odot}}$}.\\
\tablefoottext{c}{From  \cite{2013MNRAS.435.1758S}. BD: broadened disc state, HUL: hard ultraluminous state, SUL: soft ultraluminous (SUL) state.} \\
  \tablefoottext{d}{Parameter frozen at total galactic H\,I column density \citep{1990ARA&A..28..215D}.}\\
  }
 \label{tab:xmm_fit}
\end{table*}


\begin{table*}[!htbp]
 \caption{Best fit parameters for the {\it NuSTAR} observations. All errors are in the 1$\sigma$ confidence range.}
 \begin{center}
\scalebox{0.8}{   \begin{tabular}{lcccccccccc}
     \hline\hline\noalign{\smallskip}
     \multicolumn{1}{c}{Source} &
     \multicolumn{1}{c}{k${\rm T_{hot}}$} &
     \multicolumn{1}{c}{${\rm K_{hot}}^{a}$} &
     \multicolumn{1}{c}{${\rm \Gamma}$} &
     \multicolumn{1}{c}{${\rm K_{po}}^{b}$} &
     \multicolumn{1}{c}{${\rm {L/L_{edd}}^{c}}$} &
     \multicolumn{1}{r}{red. ${\chi^2/dof}$ } &
     \multicolumn{1}{c}{p } &
     \multicolumn{1}{c}{k${\rm T_{hot}}$}&
     \multicolumn{1}{r}{red. ${\chi^2/dof}$ }\\
     \noalign{\smallskip}\hline\noalign{\smallskip}
      
      \multicolumn{1}{c}{} &
      \multicolumn{1}{c}{keV} &
      \multicolumn{1}{c}{[$10^{-3}$]} &
      \multicolumn{1}{c}{} &
      \multicolumn{1}{c}{[$10^{-3}$]} & 
      \multicolumn{1}{c}{} &
      \multicolumn{1}{c}{{\tt diskbb+po}} &
      \multicolumn{1}{c}{} &
       \multicolumn{1}{c}{keV} &
      \multicolumn{1}{c}{{\tt diskpbb}} \\
      \noalign{\smallskip}\hline\noalign{\smallskip}

      \noalign{\smallskip}\hline\noalign{\smallskip}

       Ho~II X-1          & 2.11$_{-0.11}^{+0.17}$ &  5.09$_{-1.91 }^{+2.21 }$& 2.41$_{-0.26}^{+0.21}$ & 0.98$_{-0.05}^{+0.07}$ & 52.1$_{-7.36}^{+8.85}$ & 0.98/227& <0.507                   &3.31$_{-0.09}^{+0.10}$& 1.01/228 \\
       Ho~IX X-1          & 3.19$_{-0.13}^{+0.14}$ &  2.74$_{-0.46 }^{+0.56 }$& 2.38$_{-0.11}^{+0.12}$ & 2.12$_{-22.2}^{27.5 }$ & 86.9$_{-33.0}^{+40.8}$ & 1.01/440&  0.542$\pm0.009$         &4.33$_{-0.19}^{+0.21}$& 1.02/441  \\
       IC~342 X-1         & 2.37$_{-0.11}^{+0.14}$ &  9.81$_{-2,51}^{+3.15 }$ & 2.58$_{-0.18}^{+0.16}$ & 3.58$_{-1.33}^{1.60 }$ & 39.1$_{-13.3}^{+18.9}$ & 1.00/296& <0.508                   &3.41$_{-0.10}^{+0.07}$& 1.04/297  \\
       NGC~1313 X-1       & 2.64$_{-0.37}^{+0.44}$ &  2.41$_{-0.71}^{+1.61 }$ & 2.80$_{-0.23}^{+0.29}$ & 1.58$_{-0.23}^{0.29 }$ & 36.0$_{-12.3}^{+14.8}$ & 1.02/122& <0.505                   &3.52$_{-0.13}^{+0.14}$& 1.08/123  \\
       NGC~1313 X-2       & 1.94$_{-0.25}^{+0.19}$ &  1.46$_{-0.43}^{+0.63 }$ & 4.26$_{-0.22}^{+0.49}$ & 1.18$_{-0.31}^{0.84 }$ & 14.7$_{-5.00}^{+6.08}$ & 0.98/61 & <0.522                   &1.77$_{-0.07}^{+0.08}$& 0.98/62\\
       NGC~5907 ULX       & 2.51$_{-0.12}^{+0.14}$ &  2.41$_{-0.66 }^{+0.56}$ & 1.68$_{-0.41}^{+0.57}$ & 0.05$_{-0.06}^{0.04 }$ &  457$_{-46.5}^{+49.6}$ & 1.02/107&  0.579$_{-0.04}^{+0.06}$ &3.39$_{-0.07}^{+0.08}$& 1.25/108 \\
      \noalign{\smallskip}\hline\noalign{\smallskip}         
      \noalign{\smallskip}\hline\noalign{\smallskip}

      \noalign{\smallskip}\hline\noalign{\smallskip}         
          
      \noalign{\smallskip}\hline\noalign{\smallskip}
    \end{tabular}   }
 \end{center}
  \tablefoot{
  \tablefoottext{a}{Where ${\rm K_{hot}}$ is the normalisation parameter for the {\texttt {diskbb}} component. Namely ${\rm K_{hot}}$=${\rm(R_{\rm hot}/{D_{10kpc}})^{2}}\,\cos{i}$, where $R_{\rm hot}$ is the inner radius of the disk in km, ${\rm D_{10kpc}}$ is the distance in units of 10\,kpc and `i' is the inclination.  }\\
  \tablefoottext{b}{Power-law component normalisation constant: photons/keV/cm$^2$/s at 1\,keV} \\
  \tablefoottext{c}{Luminosity in the 3.-78\,keV range, extrapolated from the best-fit model. ${\rm L_{edd}}$ is the Eddington luminosity for an isotropically accreting NS with a mass of $1.4\,{\rm M_{\odot}}$}\\
  }
 \label{tab:nus_fit}
\end{table*}

\section{Discussion}
 \label{discussion}

The use of a double thermal spectrum, with similar temperatures to those observed in the dual thermal spectra of soft-state NS-XRBs, successfully describes the spectra of ULXs in our list and the unusual high-energy roll-over of the ULX spectra can be re-interpreted as the Wien tail of a hot (multicolour) black body component. The similarities between the spectral morphology of ULXs and those of NS-XRBs in the soft state are illustrated in Figure~\ref{fig:softstate}. We have plotted the \xmm spectra of two known NS-XRBs (4U 1916-05 and 4U 1705-44: see appendix~\ref{appen}) along with the spectra of two (non pulsating) ULXs from our sample. Dotted lines correspond to the dual thermal model (in this example it is an absorbed MCD plus black body  model) which --  in the case of the two NS-XRBs -- is used to model emission from the boundary layer 
and the thin accretion disk.
The same configuration is used to model the spectra of the two ULXs (in this example NGC~4559 X-1 and M81~X-6). M81~X-6 is in the BD state and NGC~4559 X-1 in the SUL state.  We stress that the unfolded spectra presented in Figure~\ref{fig:softstate}, are model dependent. They are used here in order to illustrate the apparent similarities between the spectral shapes of ULXs and soft-state NS-XRBS and not to extract any quantitative information on the spectral parameters (see also, a similar example plot in \citealt{2013MNRAS.435.1758S}). The suitability of a double thermal model for the spectra of ULXs had been noted by \cite{2006MNRAS.368..397S}; but the model was dismissed, as it was difficult to explain the presence of a secondary thermal component in terms of an accreting black hole. To probe beyond this superficial similarity, we explore the parameter space of the different spectral fits with respect to theoretical expectations, 
and discuss our findings and their implications below.

 \begin{figure}
\centering
\includegraphics[trim=0cm 0cm 0cm 0cm,  width=0.45\textwidth, angle=-0]{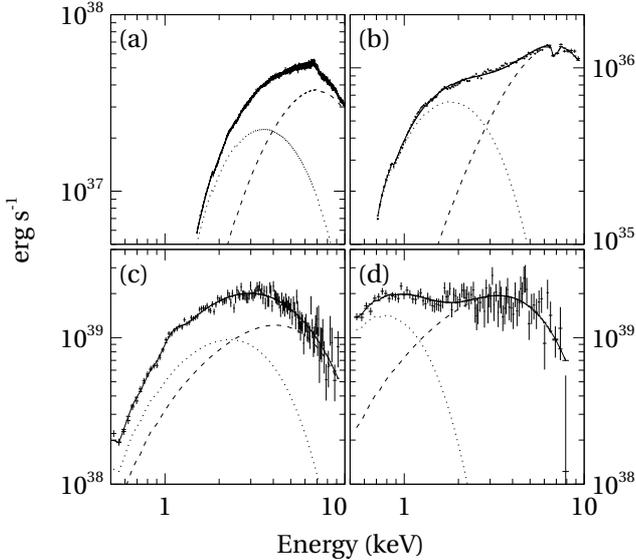}
\caption{Example, unfolded spectra of two ULXs from our sample and two well known NS-XRBs in the soft state. (a) 4U 1705-44, during a soft state. (b) Double thermal spectrum from NS-LMXB 4U 1916-05. (c) Apparent, dual thermal emission from ULX M81~X-6, at similar temperatures (see table~\ref{tab:xmm_fit}) as 4U 1705-44.  (d) Similarly shaped spectrum from NGC~4559 X-1.
 }
\label{fig:softstate}
\end{figure}

\begin{figure}
\centering
\includegraphics[trim=1.2cm 0cm 0cm 0cm,  width=0.53\textwidth, angle=-0]{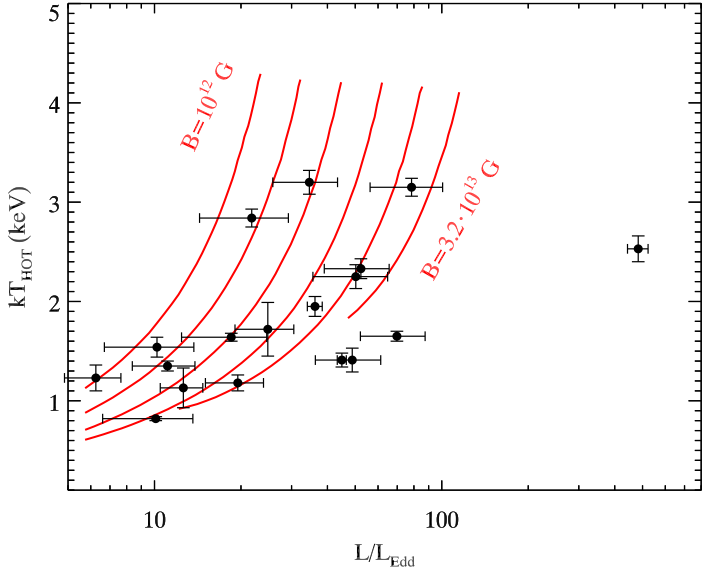}
\caption{Unabsorbed luminosity (in the 0.5-10\,keV range) vs. the temperature (in keV) of the hot multicolour disk component for the \xmm data (Table~\ref{tab:xmm_fit}). The (red) solid curves correspond to internal temperature ($T_{in}$) of the accretion curtain versus total luminosity, as predicted by \cite{2017MNRAS.tmp..143M} (see their Fig.~3).  Different curves correspond to different magnetic field strength. From left to right it is $10^{12}$, $2{\times}10^{12}$,$4{\times}10^{12}$, $8{\times}10^{12}$, $1.6{\times}10^{13}$ and $3.2{\times}10^{13}$\,G. 
 }
\label{fig:T_L}
\end{figure}

\begin{figure}
\centering
\includegraphics[trim=0cm 0cm 0cm 0cm,  width=0.5\textwidth, angle=-0]{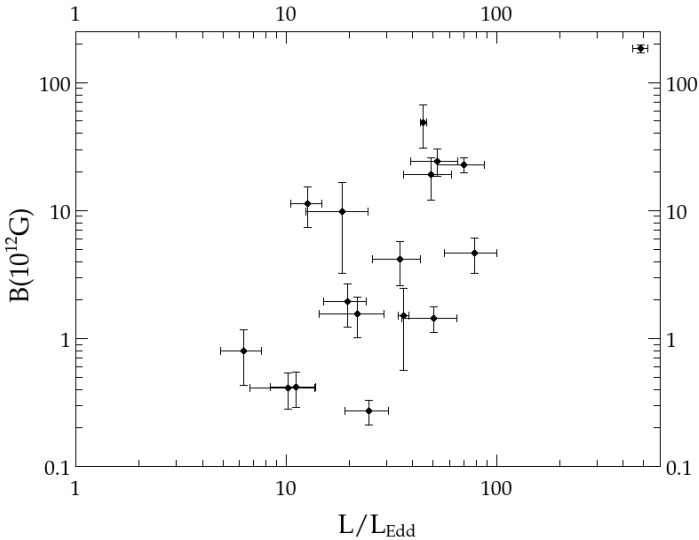}
\caption{Magnetic field strength vs.~Luminosity for the dual MCD model. All values are taken from Table~\ref{tab:xmm_fit} (columns 6 and 8). }
\label{fig:B_L}
\end{figure}

\begin{figure}
\centering
\includegraphics[trim=0cm 0cm 0cm 0cm,  width=0.5\textwidth, angle=-0]{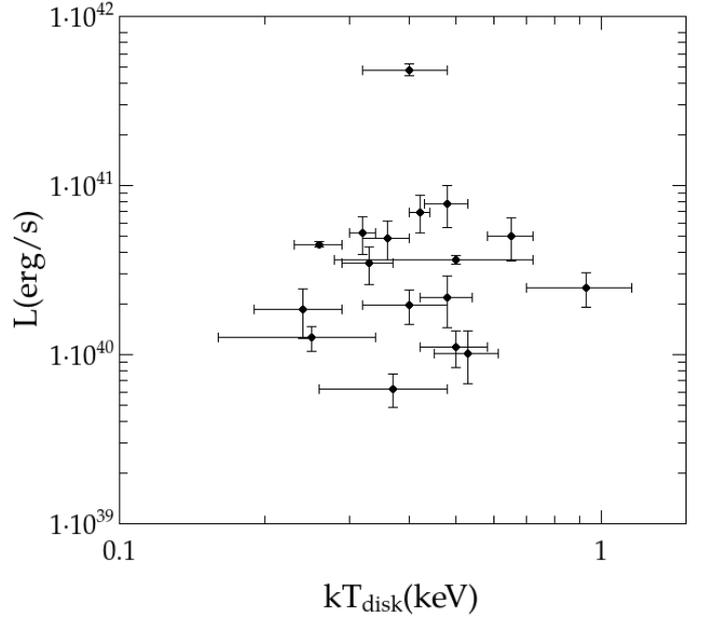}
\caption{Luminosity vs.~disk temperature for the dual MCD model. All values are taken from Table~\ref{tab:xmm_fit} (columns 3 and 6).  
 }
\label{fig:Tdisk_L}
\end{figure}

\begin{table*}[!htbp]
 \caption{Best fit parameters of the MCD plus black body model for the \xmm observations. All errors are in the 1$\sigma$ confidence range.}
 \begin{center}
\scalebox{0.8}{   \begin{tabular}{lccccccc}
     \hline\hline\noalign{\smallskip}
     \multicolumn{1}{c}{Source} &
     \multicolumn{1}{c}{nH} &
     \multicolumn{1}{c}{k${\rm T_{in}}$} &
     \multicolumn{1}{c}{${\rm R_{in}}$} &
     \multicolumn{1}{c}{k${\rm T_{bb}}$} &
     \multicolumn{1}{c}{${\rm R_{bb}}$} &
          \multicolumn{1}{c}{{${\rm R_{sph}}^{a}$}} &  
     \multicolumn{1}{r}{red. $\chi^2/dof$ }\\
     \noalign{\smallskip}\hline\noalign{\smallskip}
      
      \multicolumn{1}{c}{} &
      \multicolumn{1}{c}{[$\times10^{21}$\,cm$^2$]} &     
      \multicolumn{1}{c}{K} &
      \multicolumn{1}{c}{[km]} &
      \multicolumn{1}{c}{K} &
      \multicolumn{1}{c}{km} &
            \multicolumn{1}{c}{km} &
      \multicolumn{1}{c}{} \\
      \noalign{\smallskip}\hline\noalign{\smallskip}

      \noalign{\smallskip}\hline\noalign{\smallskip}
       Ho II X-1        & 0.40$^{b}$             & 0.57$_{-0.01}^{+0.02}$ & 724$_{-62.1}^{+76.5 }$ & 1.26$_{-0.18}^{+0.17}$ &  93.5$_{-22.7}^{+28.3}$&  974 $_{-247 }^{+297}$   & 1.25/137   \\
       Ho IX X-1        & 0.33$_{-0.08}^{+0.10}$ & 0.91$_{-0.03}^{+0.04}$ & 198$_{-42.2 }^{+38.3}$ & 1.95$_{-0.25}^{+0.27}$ &  59.2$_{-17.7}^{+18.8}$&  1120$_{-310 }^{+384}$   & 1.06/156   \\
       IC~342 X-1       & 7.42$_{-0.41}^{+0.44}$ & 0.61$\pm0.05$          & 302$_{-52.5 }^{+61.1}$ & 1.66$_{-0.10}^{+0.14}$ &  72.8$_{-19.8}^{+23.0}$&  304 $_{-104 }^{+135}$   & 1.05/101         \\
       M33 X-8          & 0.82$_{-0.02}^{+0.03}$ & 0.70$_{-0.07}^{+0.06}$ & 202$_{-41.0 }^{+51.8 }$& 1.11$_{-0.18}^{+0.16}$ &  62.1$_{-44.5}^{+50.2}$&  155 $_{-108 }^{+125}$   & 1.01/121   \\
       M81 X-6          & 1.61$_{-0.12}^{+0.16}$ & 1.03$_{-0.42}^{+0.33}$ & 155$_{-22.2 }^{+20.8 }$& 1.51$_{-0.35}^{+0.46}$ &  38.2$_{-6.78}^{+8.12}$&  346 $_{-80.1}^{+91.4}$  & 1.13/124   \\
       M83 ULX          & 0.39$^{b}$             & 0.38$_{-0.08}^{+0.10}$ & 482$_{-91.8 }^{+86.8 }$& 0.83$_{-0.11}^{+0.09}$ &  88.3$_{-16.3}^{+20.0}$&  87.2$_{-19.4}^{+22.5}$  & 0.95/44    \\
       NGC~55 ULX       & 2.16$_{-0.51}^{+0.58}$ & 0.33$_{-0.02}^{+0.03}$ & 986$_{-201 }^{+108 }$  & 0.65$_{-0.15}^{+0.16}$ &  128$_{-17.7}^{+18.3}$ &  153 $_{-29.8}^{+32.1}$  & 1.11/89    \\
       NGC~253 ULX2     & 2.35$_{-0.51}^{+0.58}$ & 0.95$_{-0.12}^{+0.14}$ & 121$_{-257 }^{+302 }$  & 1.39$_{-0.15}^{+0.16}$ &  51.8$_{-15.7}^{+14.3}$&  258 $_{-84.8}^{+105}$   & 1.05/120    \\
       NGC~253 XMM2     & 1.09$_{-0.21}^{+0.19}$ & 0.65$_{-0.10}^{+0.09}$ & 218$_{-47.6 }^{+64.3 }$& 1.17$_{-0.09}^{+0.11}$ &  54.2$_{-16.6}^{+22.0}$&  142 $_{-49.0}^{+61.4}$  & 1.14/108    \\
       NGC~1313 X-1     & 1.83$_{-0.17}^{+0.18}$ & 0.44$_{-0.20}^{+0.21}$ & 778$_{-133 }^{+156 }$  & 1.39$_{-0.43}^{+0.54}$ &  52.8$_{-22.8}^{+22.0}$&  730 $_{-186 }^{+225}$   & 1.09/113         \\
       NGC~1313 X-2     & 1.70$\pm0.11$          & 0.86$_{-0.08}^{+0.11}$ & 261$_{-35.6 }^{+43.2 }$& 1.66$_{-0.05}^{+0.06}$ &  54.6$_{-15.8}^{+19.1}$&  700 $_{-204 }^{+244}$   & 1.01/144    \\
       NGC~4190 ULX1    & 0.56$_{-0.31}^{+0.32}$ & 0.87$_{-0.38}^{+0.42}$ & 256$_{-120 }^{+208 }$  & 1.51$_{-0.19}^{+0.17}$ &  85.0$_{-6.06}^{+10.2}$&  505 $_{-30.1}^{+31.8}$  & 1.06/112   \\
       NGC~4559 X-1     & 0.15$^{b}$             & 0.43$_{-0.05}^{+0.04}$ & 1133$_{-412 }^{+528 }$ & 1.04$_{-0.10}^{+0.12}$ &  169$_ {-18.1}^{+21.5}$&  626 $_{-22.7}^{+22.2}$  & 1.07/84    \\
       NGC~4736 ULX1    & 0.86$_{-0.31}^{+0.33}$ & 0.61$_{-0.18}^{+0.22}$ & 269$_{-122 }^{+209 }$  & 0.99$\pm0.06$          &  76.6$_{-10.7}^{+11.3}$&  272 $_{-62.2}^{+73.9}$  & 1.12/68    \\
       NGC~4861 ULX     & 1.15$_{-0.81}^{+0.92}$ & 0.33$_{-0.08}^{+0.11}$ & 2398$_{-687 }^{+728 }$ & 0.93$_{-0.25}^{+0.31}$ &   205$_{-35.5}^{+44.7}$&  176 $_{-29.7}^{+35.2}$    & 1.10/22    \\
       NGC~5204 X-1     & 0.14$^{b}$             & 0.43$_{-0.12}^{+0.09}$ & 1262$_{-295 }^{+308 }$ & 0.99$_{-0.15}^{+0.18}$ &   152$_{-39.1}^{+41.0}$&  681 $_{-174 }^{+211}$   & 1.15/65    \\
       NGC~5907 ULX     & 6.04$_{-2.52}^{+3.01}$ & 0.74$\pm0.11$          & 608$_{-101}^{+133 }$   & 1.38$_{-0.09}^{+0.12}$ &   224$_{-57.7}^{+49.3}$&  6740$_{-571 }^{+554}$   & 1.09/83    \\
       NGC~7793 P13     & 0.66$\pm0.14$          & 0.63$\pm0.04$          & 251$_{-115 }^{+108 }$  & 1.70$\pm0.05$          &  59.7$_{-19.1}^{+22.0}$&  483 $_{-122 }^{+148}$   & 1.10/141 \\

      \noalign{\smallskip}\hline\noalign{\smallskip}         
      \noalign{\smallskip}\hline\noalign{\smallskip}
      \noalign{\smallskip}\hline\noalign{\smallskip}         
          
      \noalign{\smallskip}\hline\noalign{\smallskip}
    \end{tabular}   }
 \end{center}
  \tablefoot{
  \tablefoottext{a}{Spherisation radius \citep[e.g.][]{2016MNRAS.458L..10K}: ${\rm R_{sph}}=27/4{\dot m}{R_{g}}$, where  ${\rm R_{g}}=GM/c^{2}$ and ${\dot m}={\dot M}/{\dot M_{Edd}}$ .}\\
  \tablefoottext{b}{Parameter frozen at total galactic H\,I column density \citep{1990ARA&A..28..215D}.}\\
}
 \label{tab:bbody}
\end{table*}

More specifically, to investigate the case for (super-Eddington) accretion onto lowly magnetised NSs \citep[e.g.][]{2016MNRAS.458L..10K}, we applied the {\tt diskbb} plus {\tt bbody} model that is often used to model NS-XRBs in the soft state. Indeed this model describes well the spectra of the sources in our list. However, the radius inferred from the hot black body fit has a size that is approximately an order of magnitude larger\footnote{The boundary layer is expected to be a few km in size (\citealt{2009ApJ...696.1257L} and Table~\ref{appen} in this work.) } than the size of the spreading layer on the surface of the NS (see Table~\ref{tab:bbody}).  This is not surprising, since at such high accretion rates -- and for a low magnetic field NS (as considered in the \citealt{2016MNRAS.458L..10K} model) -- the accretion disk will extend well beyond the ``spherization'' radius \citep[${\rm R_{sph}}$:][]{1973A&A....24..337S}. The flow will be strongly super-Eddington and the material will, most likely, be ejected away from the surface of NS \citep[e.g.][]{2017MNRAS.468L..59K}. In this case the hot thermal component may be the result of emission of the inner disk layers, exposed by the strong outflows. In this case the hot thermal component would correspond to the stripped, inner accretion disk and the soft thermal component, as proposed in the optically thick wind scenario. Indeed, the best fit values for the size of the soft thermal component are in agreement with the ${\rm R_{sph}}$ for most of the sources in our list, alluding to the exciting possibility of accreting NSs powering a large fraction of ULXs. However, it is surprising that the dual thermal spectrum  would be almost indistinguishable between the BH-ULX and the NS-ULXs, given that in this framework the maximum temperature of the accretion disk should exceed ${\sim}4\,$keV  in the case of the NSs (perhaps even higher given the very high accretion rates of particular sources). On the other hand, there is still no strictly defined mechanism to account for the hot black body emission for the super-Eddington regime in accreting NS, and a more precise treatment may be able to resolve this apparent discrepancy. 
The fact still remains that the homogeneous fit parameters hint at the possibility of most (if not all) sources in our sample belonging to one uniform population. It is certainly plausible that this is a population of NS-XRBs instead of BH-XRBs. This implication, becomes even more intriguing when we consider the fact that two of our sources (the pulsators NGC~5907~ULX and NGC~7793~P13) are almost certainly powered by highly magnetised NSs, but -- unlike sub-Eddington NS-XRBs -- the spectra of pulsating and non-pulsating ULXs are remarkably similar.

The possibility of highly magnetised NSs powering more ULXs in our list becomes even more relevant when we consider that the most reliable and thoroughly established mechanism for sustained super-Eddington accretion episodes is the funneling of material onto the magnetic poles of high-B NSs. Indeed, in a recent publication, \cite{2017ApJ...836..113P} indicate that since the hard emission from many ULXs can be described by a combination of a hard power law and an exponential cutoff -- as expected for the emission of the accretion funnel -- this could  be considered as an indication in favour of highly magnetised NSs powering a significant fraction of ULXs. However, this claim is problematic, since -- in the presence of a high magnetic field -- the photons are expected to be concentrated in a narrow beam, most likely following the fan-beam emission diagram. Therefore, as the NS rotates, the emission should be registered in the form of characteristic pulsations. 

More importantly, the shape of their pulse profile has a complex shape comprised of two or more characteristic sharp peaks  \citep[e.g.][]{1981ApJ...251..288N, 1983ApJ...270..711W,1992hrfm.book.....M, 1996BASI...24..729P, 1997A&A...319..507P, 2004A&A...421..235R, 2013A&A...558A..74V,2014A&A...567A.129V,2016MNRAS.456.3535K}. This picture is further complicated by the fact that -- most likely -- a fraction of the fan-beam emission is scattered  by fast electrons at the edge of the accretion column and subsequently beamed towards the surface of the NS \citep{1976SvA....20..436K,1988SvAL...14..390L,2013ApJ...777..115P} off of which it is reflected, resulting in a secondary ``polar'' beam, which further complicates the pulse shape \citep[e.g.][]{2013ApJ...764...49T,kolio2017}. All but three known ULXs lack any evidence of pulsations and the three pulsating sources (NGC~5907, M82 X-2 and NGC~7793 P13) have very smooth and simple sinusoidal pulse profiles. Therefore,  very serious doubts are cast on the interpretation of the ULX spectra as being due to direct emission from the accretion column.

This contradiction appears to be resolved in a new publication by \cite{2017MNRAS.tmp..143M}. In this work, the authors demonstrate that highly magnetised NSs, accreting at high mass-accretion rates, can become engulfed in a closed and optically thick envelope (see their Fig. 1). As the primary, beamed emission of the accretion funnel is reprocessed by the optically thick material, the original pulsation information is lost.  However, if the latitudinal gradient is sufficiently pronounced -- and depending on the viewing angle and inclination of the accretion curtain -- the emission may be registered as smooth sinusoidal pulses (this could be the case of the three PULXs). More interestingly, the reprocessed emission is expected to have a multicolour black body (MCB) spectrum with a high temperature (${\gtrsim}1.0$\,keV). The hot MCB component will be accompanied by a cooler (${\lesssim}0.5$\,keV), thermal component, which originates in a truncated accretion disk. More specifically, the accretion disk is expected to extend uninterrupted, until it reaches the ${{\sim}}R_{\rm mag}$, where the material follows the magnetic field lines to form the optically thick curtain.  In this description the characteristic double thermal spectra of NS-XRBs can coexist with high-B super-Eddington accretion, thus setting NSs as excellent candidates for powering ULXs.

In light of these findings, we remodelled the \xmm spectra of the 18 ULXs, using two MCD components.  Indeed, the dual MCD model yields marginally better fits than the MCD/black-body fit, in all sources.
The best fit values for the temperature and inner radius of the cool MCD component, indicate the presence of a strong magnetic field in all the ULXs in our sample. Namely, their values are consistent with a truncated accretion disk. If we assume that the disk is truncated close to the magnetospheric radius (i.e.~to the first approximation ${\rm R_{in}}$=${\rm R_{mag}}$), we estimate that the intensity of the magnetic field exceeds $10^{12}$G in most sources in our list. 
More importantly, the temperatures of the hot MCD component and the fit-derived luminosities (see Table~\ref{tab:xmm_fit}) occupy the same parameter space, as predicted by \citeauthor{2017MNRAS.tmp..143M} (\citeyear{2017MNRAS.tmp..143M}: Their Fig. 3, and also Fig.~\ref{fig:T_L} in this work). Namely, the best-fit values for k${\rm T_{hot}}$ appear to follow the theoretical curves predicted in \cite{2017MNRAS.tmp..143M} and as a general trend, sources with stronger magnetic fields are more luminous (see Fig.~\ref{fig:B_L}) and have a hotter accretion curtain.  The observed correlation between the  magnetic field strength and the source luminosity is in agreement with the predictions of \cite{2015MNRAS.454.2539M}, where the accretion luminosity of magnetised neutron stars, in the super-critical regime, is discussed.  

Following this scheme, we  also place NGC~5907 ULX in the magnetar regime (B${\sim}1.8\,10^{14}$G) which is in agreement with the findings of \cite{2017Sci...355..817I}.  As  \citeauthor{2017Sci...355..817I} also point out, such a high value of the magnetic field is puzzling\footnote{In  \cite{2017Sci...355..817I} a multi-pole component is proposed as a resolution.}, since the source should be repeatedly entering the propeller regime \citep{1975A&A....39..185I,1986ApJ...308..669S}. However, we must underline the fact that the magnetic field values presented in this work are estimated based on the crude assumption that the $R_{\rm mag}$ is equal to the truncation radius of a thin Shakura-Sunyaev disk accreting onto a bipolar magnetic field. As such, the derived values should be treated as indications of a strongly magnetised accretor, but not considered at face value. A more realistic treatment of specific sources could yield B-field values of up to a factor of 5-6 times lower. For instance, if we re-estimate the magnetic values using the latest considerations of \cite{2017MNRAS.470.2799C} -- where it is shown that in the radiation-pressure-dominated regime, the size of the magnetosphere is independent of the mass accretion rate (see their Eqs. 39, 41 and  61) -- we end up with a value of  B${\sim}3.5\,10^{13}$G for NGC~5907 ULX and a factor of ${\sim}30-470\%$ lower magnetic field values for the other sources in our list. Nevertheless, the main outcome of our analysis remains. The best-fit parameters are consistent with our underlying assumption of a high magnetic field, which reinforces the plausibility of this scenario. A similar scenario -- in which (non pulsating) ULXs are interpreted as high-B NSs in a supercritical propeller stage -- is also proposed by \cite{2017arXiv170804502E} in a study that was submitted for publication in MNRAS, during the refereeing process of this work.

An additional implication of the high magnetic fields is the requirement that these sources are very young. Depending on the initial value of the magnetic field (which in this scenario should be at magnetar levels), the initial spin period, and the mass accretion rate, these sources are most likely younger than ${\sim}5{\times}10^{6}$\,yr, if we assume that they currently have a magnetic field of the order of $10^{12}$\,G \citep[e.g.][]{1979ApJ...234..296G,2006MNRAS.366..137Z,2016MNRAS.461....2P}. Indications in favour of a relatively young age (of the order of ${\sim}$10\,Myr) can be maintained for sources HoII~X-1, HoIX~X-1, IC342~X-1, M81~X-6, NGC~1313~X-2, NGC~253~XMM2,  NGC~253~ULX2, NGC~4559, NGC~4736 and NGC~5204,  which are associated with young stellar environments and star forming regions \citep{2005MNRAS.356...12S,2007ApJ...661..165L,2008A&A...486..151G,2008ApJ...687..471B,2010ApJ...708..354B,2011ApJ...734...23G,2013ApJ...776..100B}. Furthermore, sources  HoII~X-1, HoIX~X-1, IC342~X-1, M81~X-6, NGC~1313~X-2 and NGC~5204 have optical counterparts indicating that they are very young objects \citep{2004ApJ...603..523Z, 2004MNRAS.351L..83K,2006ApJ...641..241R}, while the nebula of IC~342~X-1 and HoIX~X-1 indicate activity of less than ${\sim}$1\,Myr \citep{2002astro.ph..2488P, 2004MNRAS.351L..83K, 2007AstBu..62...36A, 2008ApJ...675.1067F,2012ApJ...749...17C}.  However, estimation of the stellar companion's age based on the optical counterpart can be hindered by the fact that its emission may originate in the (irradiated) outer accretion disk, rather than the photosphere of the donor star \citep[e.g.][]{2012ApJ...745..123G,2012ApJ...758...85T,2012ApJ...750..110T}.  Furthermore, the magnetic field values inferred from the spectral fitting of some of our sources would require even younger ages than those derived from observations -- that is, for B${\gtrsim}10^{13}$\,G and assuming standard magnetic field decay \citep[e.g.][]{2002apa..book.....F,2006MNRAS.366..137Z}. Therefore, investigation for further indications of the presence of magnetic field in ULXs and more accurate estimation of the magnetic field strength is required to explore this intriguing scenario.

If the ``cool'' MCD component, indeed originated in a truncated accretion disk, we would expect a positive correlation between the disk temperature (${\rm T_{disk}}$) and the bolometric luminosity (L) as argued by \cite{2013ApJ...776L..36M}. In  Figure~\ref{fig:Tdisk_L} we have plotted ${\rm R_{disk}}$ versus ${\rm T_{disk}}$, however, since  the accretion disk is expected to become truncated at different values of ${\rm R_{disk}}$ (which in our scheme correspond to different B-field strength and mass accretion rates of different sources), there is a large scatter in the derived values and an accurate estimation of the $L{\sim}T$ relation cannot be attempted. 
Regarding our choice to model the cool thermal emission using a thin disk model, we must note that while in all sources analysed in this work the $R_{\rm mag}$ is larger than ${\rm R_{sph}}$ (for any plausible value of ${f_{\rm col}}$) and therefore the accretion disk could be assumed to remain thin, the $R_{\rm mag}$ values are only nominally larger than ${\rm R_{sph}}$ and -- more importantly -- the fit-inferred luminosities of the disk component are super-Eddington, suggesting that the disk will most likely be geometrically thick. In this case we would expect that the advection will perturb the thin MCD spectrum which we have used to model the cool thermal component. Nevertheless, this will not introduce any significant qualitative difference in our results (see e.g. \citealt{2013A&A...553A..61S} and also discussion in \citealt{2017MNRAS.tmp..143M}) and therefore -- as with the hot thermal component -- the {\tt diskbb} model is sufficient for the purposes of this work. Another potential issue of the geometrically thick disk is the expected emergence of outflows due to the radiation pressure, which may put the stability of this mechanism into question. However, in the presence of strong magnetic fields, the accreting, optically thick material will remain bound, even for luminosities of the order of ${\sim}10^{40}$\,erg/s \citep{2017MNRAS.tmp..143M,2017MNRAS.470.2799C}.

We also note that the numerically estimated curves in Figure~3 of \citeauthor{2017MNRAS.tmp..143M}, refer to the internal temperature (${\rm T_{in}}$: is the temperature of the inner boundary of the emission curtain, which faces the NS) of the geometrically thick accretion envelope. In the 
optically thick regime, ${\rm T_{in}}$ is related to ${\rm T_{out}}$ (which corresponds to the observed ${\rm T_{hot}}$) as,
\begin{equation}
T_{\rm in}{\approx}{T_{\rm out}}{\tau}^{\frac{1}{4}} \label{eq:tinout}
.\end{equation}
Therefore, the values of ${\rm T_{hot}}$ presented  in Figure~\ref{fig:T_L} should be multiplied by a factor of ${\approx}1.8-2.1$ (corresponding to an optically thick corona, i.e.~$\tau{\approx}10-20$) in order to represent the internal temperature (${\rm T_{in}}$). However, as stated in Section~\ref{sec-observations}, in our temperature estimations we have ignored the spectral hardening which would have produced colour-corrected temperatures given by ${\rm T_{cor}}={\rm T_{hot}}$/{\f}, with \f ranging between 1.5 and 2.1. This notable consistency between the colour correction factor and the relation between internal and external temperature in an optically thick accretion envelope further reinforces its plausibility, and with it, our confidence in the observational verification of this new scheme proposed by \cite{2017MNRAS.tmp..143M}.

These intriguing findings are also supported by the \nus observations.  Indeed, analysis of the {\it NuSTAR} data confirms the presence of the $<$10\,keV roll over, observed in the \xmm data (see e.g. Figure~\ref{fig:po}). More importantly, when the spectral curvature is modelled as hot MCB emission, the temperatures of the {\tt diskbb} component in the {\it NuSTAR} data are in agreement with the \xmm observations, particularly in those sources that were observed at similar luminosity. The presence of a thermal-like component in the \nus spectra of ULXs has also been noted for sources Circinus ULX5 \citep{2013ApJ...779..148W} and NGC~5204~X-1 \citep{2015ApJ...808...64M}, further supporting the case for emission from hot, optically thick material.

Indications for the presence of optically thick material at the boundary of the magnetosphere can also be found in sources that lie below the Eddington limit. Several X-ray pulsars (in the sub-Eddington regime) exhibit a characteristic spectral ``soft excess'', which is well described by a black body distribution at a temperature of ${{\sim}}$0.1-0.2\,keV \citep[e.g.][and references therein]{2004ApJ...614..881H}. This emission has been attributed to reprocessing of hard X-rays by optically thick material in the vicinity of the magnetosphere. The size of the reprocessing region is considerably larger than the inner edge of a standard accretion disk and it is argued that it may partially cover the primary hard emission from the accretion column \citep[e.g.][]{2006A&A...448..261Z,2006A&A...455..283L,2009A&A...494.1073R,2015MNRAS.449.3710S}. It is plausible that the predictions of \cite{2017MNRAS.tmp..143M}  are -- in essence -- an expansion of these arguments to the super-Eddington regime, where the optically thick material engulfs the entire magnetosphere, obfuscating most (or all) of the primary hard emission. The resulting accretion envelope has a temperature that is an order of magnitude higher than that of the soft excess in the sub-Eddington sources.

The \nus data also reveal the presence of a weak power law above ${\sim}15$\,keV. The hard emission, which is often present in the spectra of NS-XRBs in the soft state \citep[e.g.][and references therein]{2001AdSpR..28..307B,2007A&ARv..15....1D,2009ApJ...696.1257L}, has also been noted by \cite{2014ApJ...793...21W}, \cite{2015ApJ...806...65W}, and \cite{2017ApJ...834...77F} in \nus data of Ho~IX X-1, Ho~II X-1, and NGC~5907 ULX, respectively. The thermal emission of the accretion curtain may be modified by IC scattering from a photoionised atmosphere, analogous to an accretion disk corona \citep[e.g.][]{1980A&A....86..121S,1993ApJ...413..507H}. The presence of this highly ionised plasma is also supported by the detection of emission-like features in some of the observations in our list (see Sect.~\ref{sec-observations}). We note that the presence of similar, broad-emission-like features centred at ${\sim}1\,$keV  have also been detected in the spectra of ``nominal'' X-ray pulsars at lower accretion rates \citep[e.g.][]{2002MNRAS.337.1185R,2015MNRAS.449.3710S,2016MNRAS.458L..74L}, which also exhibit the soft excess.

We must also highlight the possibility that the apparent power-law tail may in fact be an artifact, resulting from modelling the MCB emission of the quasi-spherical accretion curtain with a MCD model. In Section~\ref{sec-observations}, it was noted that when we model the hot thermal component with a MCB model where ${\rm T_{hot}}$ is proportional to $r^{-p}$ and p is left to vary freely, the \nus spectra can be successfully fitted without the  requirement for the hard tail. More specifically, in all cases, the value of p is less than ${\sim}$0.58, which -- in the context of accretion disks -- points to an ``inflated'', advective, slim disk \citep[e.g.][and references therein]{2004ApJ...601..428K}. In this case, it is fairly plausible that the radically different geometry of the accretion curtain will cause a significant deviation from the  temperature gradient of $T{{\sim}}r^{-0.75}$ assumed by the standard MCD model used in our fits, resulting in an underestimation of the hard emission, which appears as an excess above ${\sim}15\,$keV. 

The hypothesis of an ``obscured'', highly magnetised NS as the central engine in ULXs may also resolve the contradiction regarding the only known (to this date) detection of a relativistic jet in a ULX, in Ho~II~X-1 \citep{2015MNRAS.452...24C}. While collimated jets are often detected in BH-XRBs and AGN, they are only present during the low-accretion, non-thermal {\it hard state}. In the case of Ho~II~X-1, though, the collimated jet is detected in a high accretion state, during which the spectrum is dominated by thermal emission, which is in stark contrast to the BH-XRB/jet paradigm. This contradiction is resolved when we consider a high-B NS as the accretor, which -- for sufficiently high values of the magnetic field -- can power collimated relativistic jets at high accretion rates \citep{2017MNRAS.469.3656P,2016ApJ...822...33P}.
However, in this framework the presence of the jet is also contingent upon the NS spin period. Only a limited set of parameters would yield a powerful jet in the high accretion-rate regime, which may explain the lack of a jet in most known ULXs.  It is also certainly plausible that the non-detection of radio jets may be the result of a lack of sensitivity, since the expected flux would most likely lie in the few $\mu$\,Jy range or less. Given the above discussion and the bulk of theoretical expectations, strong outflows should be expected for any type of accretor (i.e. BH, highly or lowly magnetised NS). The more pertinent question, then, would be whether the soft thermal component originates in a hot optically thick wind component, close to the accretor or the truncated accretion disk. Therefore, given the recent considerations regarding the different candidates for the ULX accretors, it is important that the evolution of the $L_{soft}{{\sim}}{T_{in}}$ relation for specific sources is revisited.

The ``universal'', power-law-shaped luminosity function of ULXs and HMXBs \citep[e.g.][]{2004NuPhS.132..369G,2004ApJS..154..519S,2012MNRAS.419.2095M} may also be interpreted as favouring NS-powered ULXs. More specifically, the smooth shape of the HMXB luminosity function up to ${\log L}{\sim}40.5$ strongly implies that ULXs are composed of ordinary HXMBs with stellar-mass accretors. Since most HMXBs are powered by NSs \citep[e.g.][]{2006A&A...455.1165L,2009ApJ...707..870B}  and also most ULXs are found in star-forming regions \cite[e.g.][and references therein]{2011NewAR..55..166F} that favour the evolution of NS-HMXBs, it is reasonable to postulate that most ULXs are indeed NS-XRBs. It is also of great interest to investigate if there are fundamentally different characteristics between ULXs and sources that lie above the ${\sim}10^{40}$\,erg/s break in the luminosity function \citep{2012MNRAS.419.2095M}. Indeed the two brightest HLXs -- M82~X-1 and ESO~243-49~HLX-1 -- do not feature the spectral cutoff of ULXs \citep[e.g.][]{2006ApJ...637L..21D,2009Natur.460...73F} and also appear to transition between the empirical BH-XRB accretion states  (e.g. \citealt{2009ApJ...705L.109G}; \citealt{2010ApJ...712L.169F}, although \citealt{2016ApJ...829...28B} recently indicated that M82~X-1, during episodes of high accretion, can be modelled as a stellar-mass BH, accreting at super-Eddington rates). The differing aspects between sources above and below the luminosity break indicate a different type of accretor between HLXs and ULXs. Within the scheme discussed in this work, this could mean that while most ULXs have NS accretors, HLXs harbour either supercritically accreting stellar-BHs or sub-Eddington accreting IMBHs. Nevertheless, given the very small sample of HLX sources, such hypotheses remain strictly in the realm of speculation.

\section{Conclusion}
\label{conclusion}   
We have presented an alternative interpretation of the X-ray spectra of eighteen well-known ULXs, which provides physically meaningful spectral parameters. 
More specifically, from the analysis of the \xmm and \nus spectra, we  note that the curvature above ${{\sim}}$5\,keV -- found in the spectra of most ULXs -- is consistent with the Wien tail of thermal emission in the $>$1\,keV range. Furthermore the high-quality \xmm spectra confirm the presence of a secondary, cooler thermal component.  These findings are in agreement with the analysis presented in previous works \citep[e.g.][]{2014ApJ...793...21W,2015ApJ...806...65W,2016MNRAS.460.4417L}. However, in contrast to the currently accepted paradigm, we propose that the dual thermal spectrum may be the result of accretion onto a highly magnetised NS, as predicted in recent theoretical models \citep{2017MNRAS.tmp..143M} in which the hot thermal component originates in an optically thick envelope that engulfs the entire NS at the boundary of the magnetosphere, and the soft thermal component originates in an accretion disk that becomes truncated at approximately the magnetospheric radius. We claim that this finding offers an additional and compelling argument in favour of neutron stars as more suitable candidates for powering ULXs, as has been recently suggested \citep{2016MNRAS.458L..10K,2017MNRAS.468L..59K}. In light of this interpretation, the ultraluminous state classification put forward by \cite{2013MNRAS.435.1758S} can be re-interpreted in terms of different temperatures and relative flux contribution of the two thermal components, which result in the different spectral morphologies.

Nevertheless, we stress that there is considerable degeneracy between different models that can fit the spectra equally well, and so far there are no observational features, such as cyclotron lines or transitions to the propeller regime \citep[e.g.][]{1972A&A....21....1P,1973ApJ...184..271L,1975A&A....39..185I}, that will conclusively favour this hypothesis over other comprehensive and equally plausible interpretations (i.e.~optically thick outflows from critically accreting black holes). Furthermore, the presence of strong outflows is also expected in the case of accreting high field NSs,  which may account for the soft thermal emission in the NS scenario as well. Given the encouraging results of this work, further examination of this scenario is warranted. To this end, the fractional variability of ULXs  (which is addressed by the BH super-Eddington wind model) should be reviewed in the context of the highly magnetised NS model, and the possibility  of aperiodic flux variation due to the rotating accretion curtain must be explored further. Moreover, deeper broadband observations that will also allow precise phase resolved spectroscopy of the pulsating sources, as well as long-term monitoring -- in sources for which this is feasible -- are necessary in order to further probe this newly emerging paradigm.

\section{Acknowledgements}
The authors would like to thank the anonymous referee whose contribution significantly improved our manuscript. Also, F.~K., O.~G., N.~W. and D.~B. acknowledge support from the CNES. F.~K. warmly thanks Apostolos Mastichiadis and Maria Petropoulou for comments and stimulating discussion.

\begin{appendix}
\section{Spectral extraction and analysis of the NS-XRBs}

\xmm spectra for NS-XRBs 4U 1705-44 and 4U 1916-05 were extracted and analysed using the same procedures as described in Section~\ref{sec-observations}. Both sources where fitted with an absorbed MCD plus black body model ({\tt xpsec} model {\tt tbnew\_gas(diskbb+bbody)}), where the MCD model was used for the ``cool'' thermal emission of the truncated accretion disk and the black body for the ``hot'' thermal emission, expected to originate in the boundary layer formed on the surface of the NS. Best fit values are presented in Table~\ref{appen}. The inner disk radius inferred from the normalisation parameter of the {\tt diskbb} component was estimated using the expression given in Table~\ref{tab:xmm_fit}. We assumed an inclination of $i=60\deg$ for 4U 1705-44 and $i=80\deg$ for 4U 1916-05 (this is an edge-on viewed source \citealt{2004A&A...418.1061B}). The size of the black body emitting region (column 6) was estimated using the expression given in Table~\ref{tab:bbody}. Distances of 8\,kpc and 9\,kpc were assumed for 4U 1705-44 and 4U 1916-05, respectively.

\begin{table*}[!htbp]
 \begin{center}
  \begin{threeparttable}
    \caption{ Best fit values for \xmm observations 0085290301 and 0551270201 of sources 4U 1916-05 and 4U 1705-44, respectively.}
     \begin{tabular}{lccccc}
        \toprule
        Source      & nH                        &  k${\rm T_{in}}$    &${\rm R_{in}}$           &${\rm T_{BB}}$ &${\rm R_{BB}}$   \\
                    &                           &                     &                         &               &              \\
                    &  $\times10^{21}$\,cm$^2$  & keV                 &  km                     & keV           &     km         \\
        \midrule
         4U 1705-44 & 22.5$\pm0.13$             & 1.14$\pm0.03$       &  11.8$_{-0.38}^{+0.37}$ & 1.76$\pm0.01$ & 5.36$_{-0.19}^{+0.22}$     \\
         4U 1916-05 & 2.15$\pm0.03$             & 0.66$\pm0.01$       &  13.2$_{-0.12}^{+0.31}$ & 1.75$\pm0.01$ & 1.23$_{-0.01}^{+0.02}$             \\

        \bottomrule
     \end{tabular}
    \begin{tablenotes}
      \small
      \item \footnotesize{}
    \end{tablenotes}
      \label{appen}
  \end{threeparttable}
\end{center}

\end{table*}

\end{appendix}
\bibliography{/home/filippos/Documents/Bibliography/general}

\end{document}